\documentclass[prd,aps]{revtex4}
\usepackage{color}
\usepackage{mathtools}
\usepackage{array}
\usepackage{tabularx}
\newcolumntype{Y}{>{\raggedright\arraybackslash}X}
\usepackage{subcaption}
\usepackage{diagbox}
\usepackage{array}
\usepackage{tikz}
\usetikzlibrary{decorations.pathmorphing}
\usepackage{bm}
\usepackage{amsmath,amssymb,amsfonts,dcolumn,graphicx,latexsym,epsfig}
\usepackage{bbold}
\usepackage{rsfso}
\usepackage{microtype}
\usepackage{setspace} 
\usepackage{array, booktabs}
\usepackage[utf8]{inputenc}
\UseRawInputEncoding
\usepackage{hyperref}
\hypersetup{
colorlinks=true,
linkcolor=blue,
filecolor=magenta,      
urlcolor=magenta,
citecolor=magenta,
}
\usepackage{multirow}
\usepackage{natbib}
\usepackage{float}
\usepackage{booktabs}
\usepackage{pifont}
\usepackage{mathrsfs}
\usepackage{caption}
\begin{document}

\title{Signatures of charged rotating regular black holes: quasinormal modes, grey-body factors and shadows}
\author{Abhisek Barman Maji}
\email[Email address: ]{rajyavisek027@kgpian.iitkgp.ac.in}
\affiliation{Department of Physics, Indian Institute of Technology, Kharagpur 721 302, India}

%\author{Sayan Kar}
%\email[Email address: ]{sayan@phy.iitkgp.ac.in}
%\affiliation{Department of Physics, Indian Institute of Technology, Kharagpur 721 302, India}

\begin{abstract}

\noindent We construct and study possible astrophysical signatures of the rotating counterparts of two charged regular black holes, namely the Ay\'on--Beato--Garc\'ia (ABG) and Balart--Panotopoulos--Rinc\'on (BPR) spacetimes -- obtained using the Newman-Janis algorithm. After verifying the regularity of the rotating metrics, we study massless scalar perturbations, compute the quasinormal mode (QNM) spectra, and characterise their dependence on the spin and charge parameters. Grey-body factors (GBFs) are computed, and the QNM-GBF correspondence is verified. The low-frequency superradiant amplification factor is also examined via matched asymptotic expansions. The Lyapunov exponents are used to connect the eikonal QNM damping rate to photon-orbit instability. Additionally, from shadow profiles, we constrain the parameter space of rotating regular black holes using the Event Horizon Telescope (EHT) observations. Across all the probes, we find that the rotating BPR and Kerr--Newman black holes behave almost indistinguishably from one another, while the rotating ABG geometry exhibits distinct perturbative, scattering, and geometric properties, particularly at larger values of spin and charge. Our analysis indicates that differentiating between charged rotating regular black hole models may be possible using gravitational wave and shadow observations.

\end{abstract}


\maketitle

\section{Introduction}

\noindent The classical black hole solutions of general relativity -- Schwarzschild, Reissner--Nordstr\"om, and Kerr, all contain a curvature singularity hidden behind an event horizon. The singularity theorems of Hawking and Penrose show that the formation of such singularities is a generic feature of general relativistic solutions, provided reasonable energy conditions and causality assumptions hold \cite{Hawking1970, Penrose1965, Hawking1966, Senovilla2015}. This has motivated a long-standing effort to construct black hole models in which the central singularity is removed while the essential features of the classical solutions, like an event horizon, asymptotic flatness, and the associated causal structure, are preserved.\\

\noindent Regular black holes realise this idea by replacing the classical singularity with a finite-curvature core, so that all curvature invariants remain finite throughout the spacetime and geodesics can, in many cases, be extended smoothly through the central region. The first such construction was proposed by Bardeen \cite{Bardeen1968}, who introduced a static, spherically symmetric geometry with a de Sitter core and an asymptotically Schwarzschild exterior. Ay\'on-Beato and Garc\'ia later showed that the Bardeen-type metric, and a related charged generalisation, can be obtained as an exact solution of Einstein gravity minimally coupled to nonlinear electrodynamics (NLED) \cite{AyonBeato2000, AynBeato1998, AynBeato1999}, placing regular black holes on a firmer field-theoretic footing rather than treating them as purely phenomenological metrics. Since then, a variety of regular geometries have been proposed from different theoretical starting points, including further NLED-sourced solutions \cite{Hayward2006, Bronnikov2001, Uchikata2012, Balart2014, Dymnikova2015, PoncedeLeon2017, Kar2023dko, Kar2024a}, solutions with quantum-gravity inspired corrections, and modifications thereof \cite{BenAchour2018khr, Calza2024xdh, Konoplya2023aph, Kobayashi2016gtl}, geometries constructed using modified gravity \cite{Berej2006, Pinto2025loq, Ma2017jko}, dark matter profiles \cite{Konoplya2025ect, Kar2025phe} and many more \cite{Dymnikova1992, Roman1983, Frolov2017, Simpson2020, Borissova2025, Bronnikov2024, Bueno2025}. The characteristics and observable signatures of regular black holes have been extensively studied \cite{Nomura2005dn, Kumar2020a, Schee2015, CarballoRubio2018, Kumar2025our, Bhattacharyya2026, Agrawal2026rwu}. \\

\noindent In this work, we focus on two such models with asymptotic mass $M$ and charge $q$ which reduce to the Reissner--Nordstr\"om geometry for $r \gg q$ and to Schwarzschild as $q \to 0$: the Ay\'on-Beato--Garc\'ia (ABG) solution \cite{AynBeato1998}, and the Balart--Panotopoulos--Rinc\'on (BPR) solution \cite{Balart2023}. We refer to these asymptotically Reissner--Nordström geometries as charged regular black holes \cite{Balart2023}. Since astrophysical compact objects generically carry angular momentum, a static regular black hole is, by itself, of limited direct relevance to observations; a rotating generalisation is needed. The Newman-Janis algorithm, originally introduced to derive the Kerr metric from the Schwarzschild solution \cite{Newman1965}, together with its later refinements and complexification prescriptions \cite{Bambi2013}, has been used for generating rotating counterparts of static regular black holes \cite{Ghosh2020ece, Bambi2013, Neves2014aba, Toshmatov2014, Toshmatov2017, Ghosh2015, Jusufi2019caq}. While the resulting metrics are not, in general, guaranteed to solve the Einstein equations for a well-defined matter source, they reduce to Kerr in the appropriate limits and retain its qualitative structure, while replacing the ring singularity with a regular core. Such solutions have consequently been widely used as theoretical probes for exploring how the regularisation manifests in astrophysical quantities, including particle motion, energy extraction \cite{Ghosh2014, Mohamed2024, Shahzadi2018, Kar2025a}, gravitational lensing, black hole shadows \cite{Abdujabbarov2016, Amir2016, Tsukamoto2017fxq, Jusufi2018, Kumar2020, Salehi2024, Belhaj2023, Ramadhan2023, Ban2026}, and the quasinormal modes (QNMs) corresponding to a perturbation \cite{Jusufi2020, Franzin2022, Yang2026, Li2023, Pedrotti2024, Khoo2025, Peng2025}.\\

\noindent This raises the central question of this article: to what extent do these different regularised spacetimes remain observationally distinguishable from one another, and from Kerr--Newman? Answering this requires examining several complementary characteristics, such as the QNM spectrum, the photon sphere, the shadow boundary, and the scattering properties encoded in grey-body factors (GBFs) and superradiance. Each of these probes a different aspect of the geometry, from the near-horizon structure to the asymptotic wave-scattering behaviour. By examining these probes, we aim to determine whether there is a clear distinction between regular and singular solutions and to explore whether different regularisations are sufficient to guarantee mutual observational degeneracy. The study of these signatures has gained significant observational relevance over the last decade or so through the emergence of a large amount of data and results in gravitational wave astronomy (LIGO-Virgo-KAGRA collaboration \cite{PhysRevX.13.041039, LIGOScientific2025slb, LIGOScientific2026wfs}) and infrared imaging of astrophysical objects (the Event Horizon Telescope (EHT) \cite{M872019, EventHorizonTelescopeCollaboration2022}).\\

\noindent With this aim in mind, we probe these geometries with the same suite of diagnostics. First, we map the horizon structure and ergo-region in the $(a,q)$ parameter space, verify curvature regularity using a minimal set of Weyl, Ricci, and mixed invariants, and examine the weak energy condition. The QNM frequencies corresponding to scalar perturbations are computed using both the spectral method and the sixth-order Wentzel--Kramers--Brillouin (WKB) approximation, and their dependence on spin and charge parameters is examined. Next, we address the scattering problem with GBFs along with low-frequency superradiance via matched asymptotic expansion. We use the Hamilton--Jacobi formalism to obtain shadow silhouettes and place bounds on the spin ($a$) and charge ($q$) parameters of rotating regular spacetimes using observational results from the EHT collaboration. We also study the Lyapunov exponents governing the eikonal-limit QNM correspondence.\\

\noindent This article is organised as follows. In section \ref{geometry}, we construct rotating generalisations of the ABG and BPR regular black holes via the Newman-Janis algorithm and discuss their properties. We then obtain the scalar perturbation equation and compute the scalar QNM spectra in section \ref{QNM}. In section \ref{Grey}, we compute GBFs and their correspondence with the QNM spectrum as well as the superradiant amplification factor, followed by section \ref{photon}, where we obtain the photon sphere and shadow together with the Lyapunov exponent. Finally, we conclude in section \ref{conclusion} with comments on future directions. We examine a few more charged regular black holes and compare the QNM spectra of their rotating counterparts with that of the Kerr--Newman black hole in Appendix \ref{appendix}. In Appendix \ref{methods}, we briefly discuss the two methods employed in this article to compute the QNMs. The Natural units ($G=c=1$) are used throughout this article.

\section{Family of charged rotating regular black holes} \label{geometry}

\noindent Regular black holes provide non-singular alternatives to the classical general relativistic (GR) solutions, replacing the central curvature singularity with a finite-curvature core while retaining an event horizon. We begin with a general static regular black hole geometry with the line element
\begin{equation}
    ds^2=-\left(1-\frac{2M(r)}{r}\right)dt^2+\left(1-\frac{2M(r)}{r}\right)^{-1}dr^2+r^{2}\Big(d\theta^{2}+\sin^{2}\theta d\phi^{2}\Big), \label{reg}
\end{equation}
where $M(r)$ is the mass function that depends on the asymptotic mass and charge of the black hole. In this article, we will focus on a regular black hole constructed by Ay\'on-Beato and Garc\'ia \cite{AynBeato1998}. It is an exact solution of Einstein gravity coupled to nonlinear electrodynamics (NLED) with the line element
\begin{equation}
    ds^{2} = -\Big(1-\frac{2Mr^{2}}{(r^{2}+q^{2})^{3/2}}+\frac{q^{2}r^{2}}{(r^{2}+q^{2})^{2}}\Big)dt^{2} + \Big(1-\frac{2Mr^{2}}{(r^{2}+q^{2})^{3/2}}+\frac{q^{2}r^{2}}{(r^{2}+q^{2})^{2}}\Big)^{-1}dr^{2}  +r^{2}\Big(d\theta^{2}+\sin^{2}\theta d\phi^{2}\Big). \label{ABG}
\end{equation}
Here, $M$ is the ADM mass of the spacetime and $q$ is the NLED charge parameter (electric or magnetic). This static black hole solution satisfies the weak energy condition, and in the weak-field limit, the nonlinear electromagnetic field that sources it reduces to the Maxwell field. Along with the curvature invariants, the electric field ($\mathcal{E}$) associated with this metric is also regular everywhere and reduces to the familiar Coulombic form $\mathcal{E} \sim q/r^2$ at the large - $r$ limit \cite{AynBeato1998}. We will refer to this geometry as the ABG regular black hole.\\

\noindent We will also study another charged regular black hole geometry with asymptotic mass $M$ and charge $q$. The metric was constructed by Balart, Panotopoulos and Rinc\'on \cite{Balart2023}, with mass function
\begin{equation}
    M(r)=M\left(\exp\left(-\frac{q^2}{2Mr}\right)-\frac{q^3r^4}{2M(r^2+q^2)^3}\right). \label{BPR}
\end{equation}
This is a modification of one of the regular black hole solutions presented in \cite{Balart2014}. From now on, it will be referred to as the BPR black hole. Both the spacetimes mentioned above approach Reissner--Nordstr\"om geometry at the asymptotic limit ($r \gg q$) and approach the Schwarzschild solution in the $q\rightarrow0$ limit.\\

\noindent Since astrophysical black holes are expected to possess non-zero angular momentum, it is crucial to study the signatures of rotating regular black holes. To construct the rotating counterparts of the charged regular black holes, we will be using the Newman-Janis algorithm with type I complexification \cite{Bambi2013}.\\

\noindent We start with Eq.~\eqref{reg} and use the Newman-Janis algorithm \cite{Newman1965} to obtain the rotating black hole geometry
\begin{equation}
\begin{split}
    \text{d}s^2 &= -\left(1- \frac{2M(r) r}{\Sigma}\right)\text{d}t^2- \frac{4M(r) r a\sin^2\theta}{\Sigma} \text{d}t\text{d}\phi + \frac{\Sigma}{\Delta} \text{d}r^2 + \Sigma \text{d}\theta^2 \\
    &+ \sin^2\theta \left( r^2 + a^2 + \frac{2M(r) r a^2\sin^2\theta}{\Sigma} \right)\text{d}\phi^2,
    \end{split} \label{regrot}
\end{equation}
where $\Sigma=r^2+a^2 \cos^2{\theta}$, and $\Delta=r^2-2M(r)r+a^2$. The line element in Eq.~\eqref{regrot} is a type I solution as it was obtained with complexification of type I \cite{Bambi2013} where we complexify the $1/r$ term, without altering the mass term $M(r)$.\\

\begin{figure}[H]
    \centering

    \begin{subfigure}{0.32\textwidth}
        \centering
        \includegraphics[width=\linewidth]{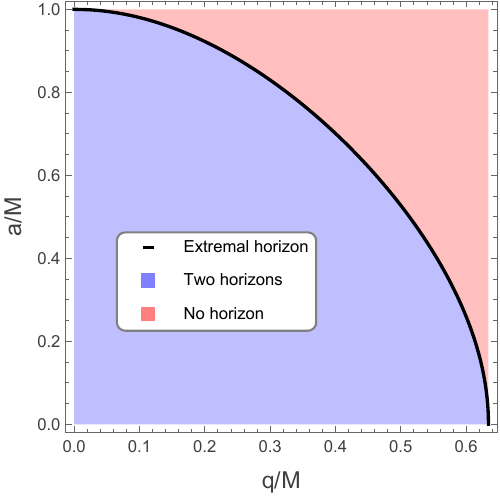}
    \end{subfigure}
    \hfill
    \begin{subfigure}{0.32\textwidth}
        \centering
        \includegraphics[width=\linewidth]{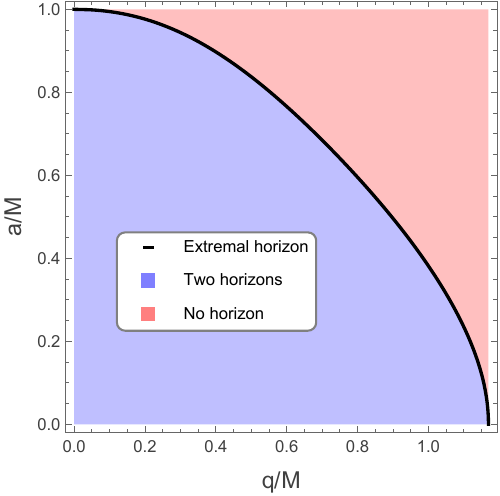}
    \end{subfigure}
    \hfill
    \begin{subfigure}{0.32\textwidth}
        \centering
        \includegraphics[width=\linewidth]{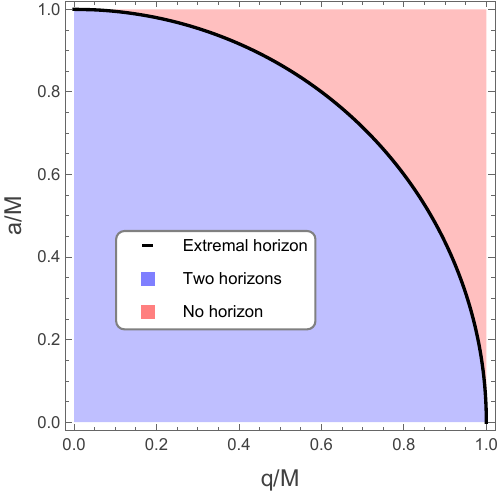}
    \end{subfigure}

    \caption{Parameter space for black holes with two, one and no horizon. (left) rotating ABG black hole. (middle) rotating BPR black hole. (right) Kerr--Newman black hole.}
    \label{fig:hor}
\end{figure}

\noindent These rotating geometries possess null hypersurfaces (horizons), determined by the condition $\Delta(r)=0$. The horizon occurs at a constant radial distance and is independent of the $\theta$ coordinate. Depending on the values of $M$, $a$ and $q$, the $\Delta(r)=0$ equation can have two positive roots corresponding to the outer event horizon and inner Cauchy horizon. At the extremal configuration, these horizons coincide, and for sufficiently large values of $a$ and $q$, the horizons can disappear, leaving a regular horizonless spacetime. In Fig.~\ref{fig:hor}, we have shown how the existence and number of horizons depend on the rotation parameter and the charge parameter. The extremal values of the charge parameter for rotating ABG and rotating BPR spacetimes are $0.6342M$ and $1.1694M$, respectively.

\begin{figure}[H]
    \centering

    % Row 1
    \begin{subfigure}{0.24\textwidth}
        \centering
        \includegraphics[width=\textwidth]{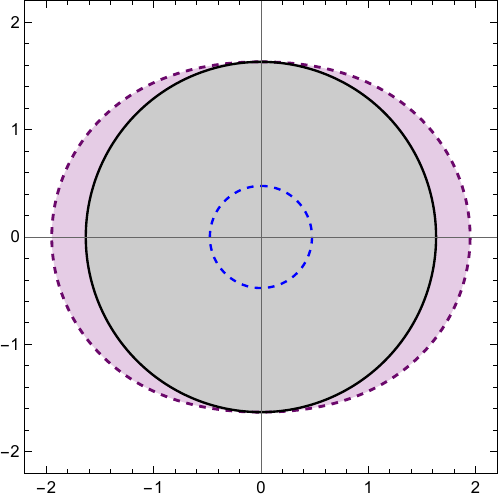}
        \caption{\footnotesize ABG ($a=0.7M,q=0.2M$)}
    \end{subfigure}
    \hfill
    \begin{subfigure}{0.24\textwidth}
        \centering
        \includegraphics[width=\textwidth]{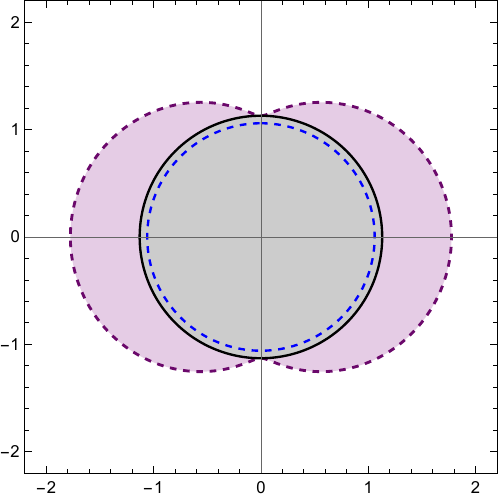}
        \caption{\footnotesize ABG ($a=0.7M,q=0.4M$)}
    \end{subfigure}
    \hfill
    \begin{subfigure}{0.24\textwidth}
        \centering
        \includegraphics[width=\textwidth]{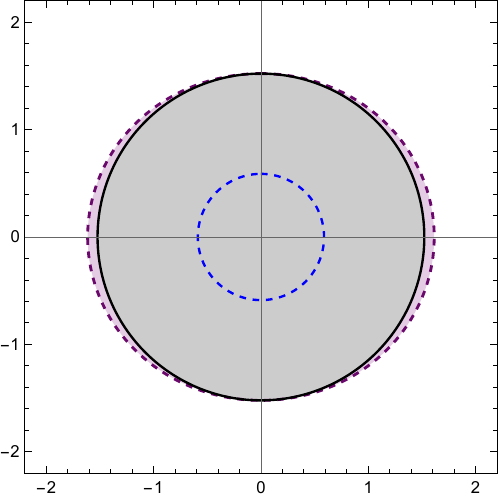}
        \caption{\footnotesize ABG ($a=0.2M,q=0.5M$)}
    \end{subfigure}
    \hfill
    \begin{subfigure}{0.24\textwidth}
        \centering
        \includegraphics[width=\textwidth]{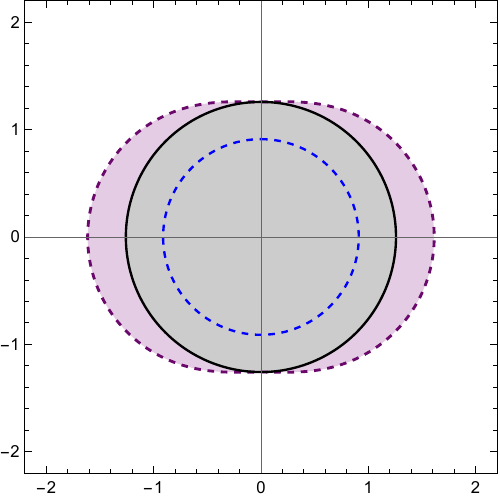}
        \caption{\footnotesize ABG ($a=0.5M,q=0.5M$)}
    \end{subfigure}

    \vspace{0.5cm}

    % Row 1
    \begin{subfigure}{0.24\textwidth}
        \centering
        \includegraphics[width=\textwidth]{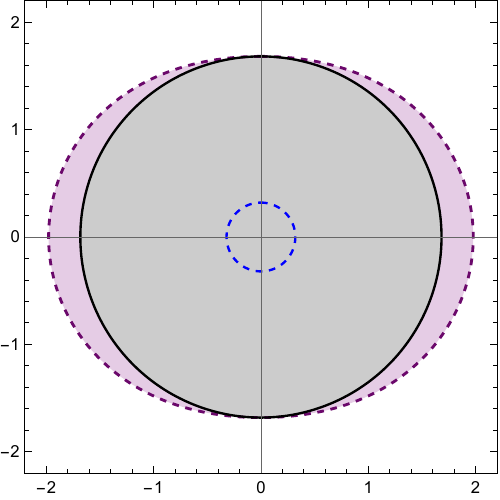}
        \caption{\footnotesize BPR ($a=0.7M,q=0.2M$)}
    \end{subfigure}
    \hfill
    \begin{subfigure}{0.24\textwidth}
        \centering
        \includegraphics[width=\textwidth]{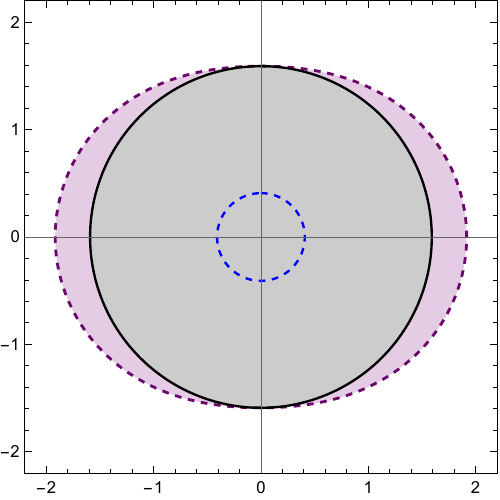}
        \caption{\footnotesize BPR ($a=0.7M,q=0.4M$)}
    \end{subfigure}
    \hfill
    \begin{subfigure}{0.24\textwidth}
        \centering
        \includegraphics[width=\textwidth]{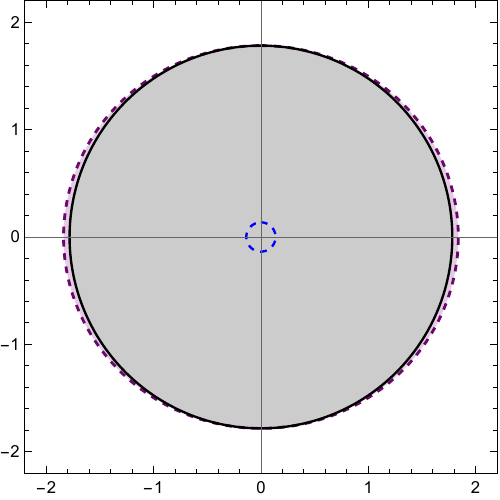}
        \caption{\footnotesize BPR ($a=0.2M,q=0.5M$)}
    \end{subfigure}
    \hfill
    \begin{subfigure}{0.24\textwidth}
        \centering
        \includegraphics[width=\textwidth]{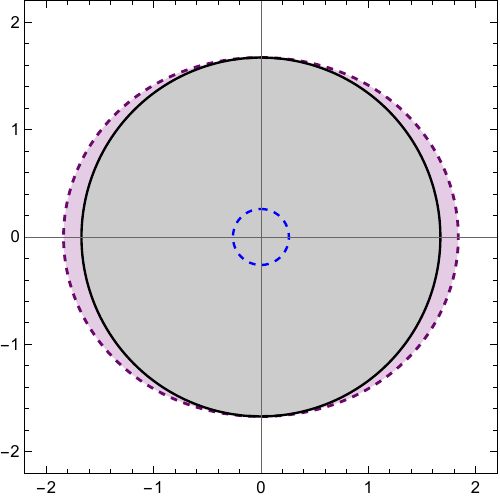}
        \caption{\footnotesize BPR ($a=0.5M,q=0.5M$)}
    \end{subfigure}

\caption{Outer event horizon (solid black), inner horizon (dashed blue), static limit surface (dashed purple) and ergo-region (purple shaded region) for the rotating ABG (top row) and rotating BPR (bottom row) black hole corresponding to different values of $a$ and $q$.}
\label{fig:ergo}
\end{figure}

\noindent In rotating spacetimes, we also encounter the frame-dragging effect, in which the rotation of the black hole causes the dragging of local inertial frames. This leads to the formation of the ergo-region enclosed between the outer event horizon and the static limit surface (or infinite redshift surface), where a particle cannot remain static with respect to an observer at infinity and is forced to co-rotate with the black hole. The outer boundary of the ergo-region, the so-called static limit surface, is determined by the condition $g_{tt}=0$ and coincides with the event horizon along the rotation axis. Parametric dependence of the event horizon, inner horizon, static limit surface, and ergo-region for the rotating ABG and rotating BPR regular black hole is shown in Fig.~\ref{fig:ergo}. The ergo-region grows with both the rotation parameter $a$ and charge parameter $q$.

% \subsubsection{Curvature scalars}

\begin{figure}[H]
    \centering

    % First row
    \begin{subfigure}{0.31\textwidth}
        \centering
        \includegraphics[width=\linewidth]{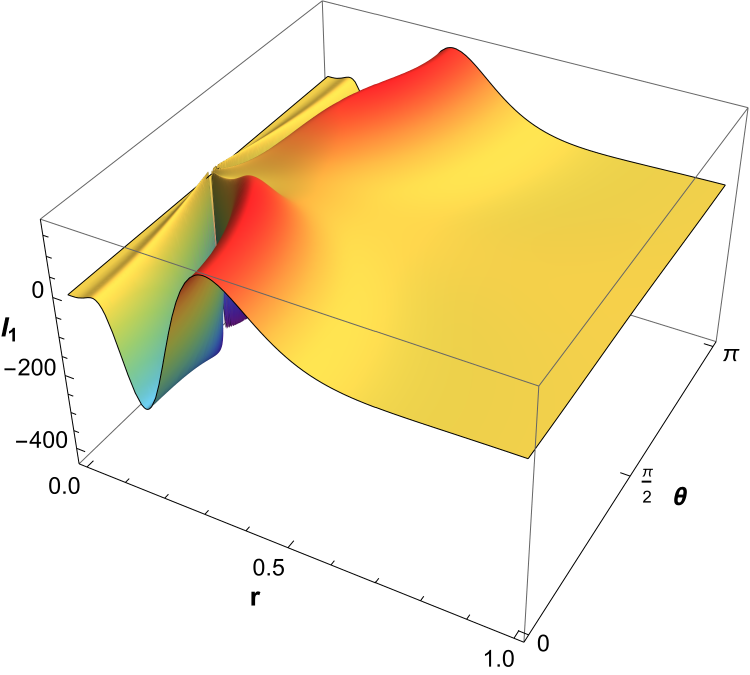}
    \end{subfigure}
    \hfill
    \begin{subfigure}{0.31\textwidth}
        \centering
        \includegraphics[width=\linewidth]{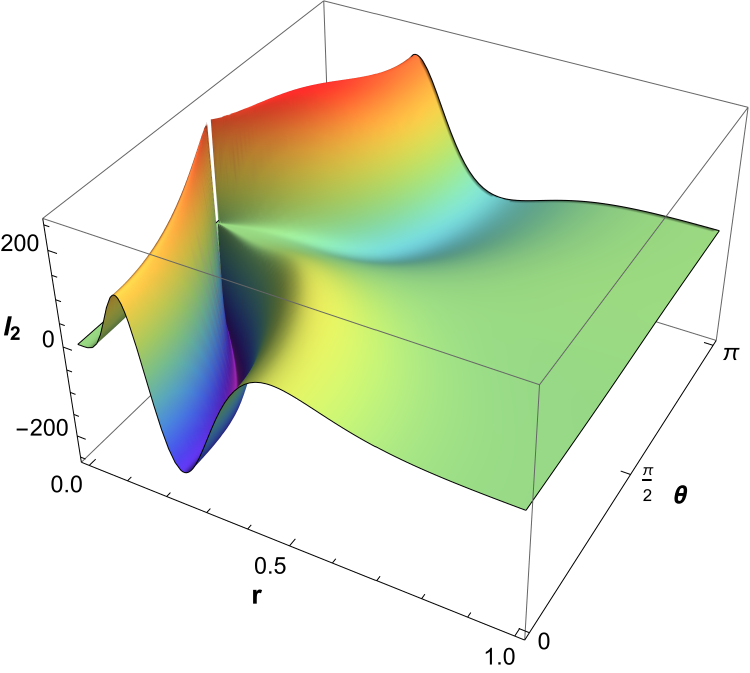}
    \end{subfigure}
    \hfill
    \begin{subfigure}{0.31\textwidth}
        \centering
        \includegraphics[width=\linewidth]{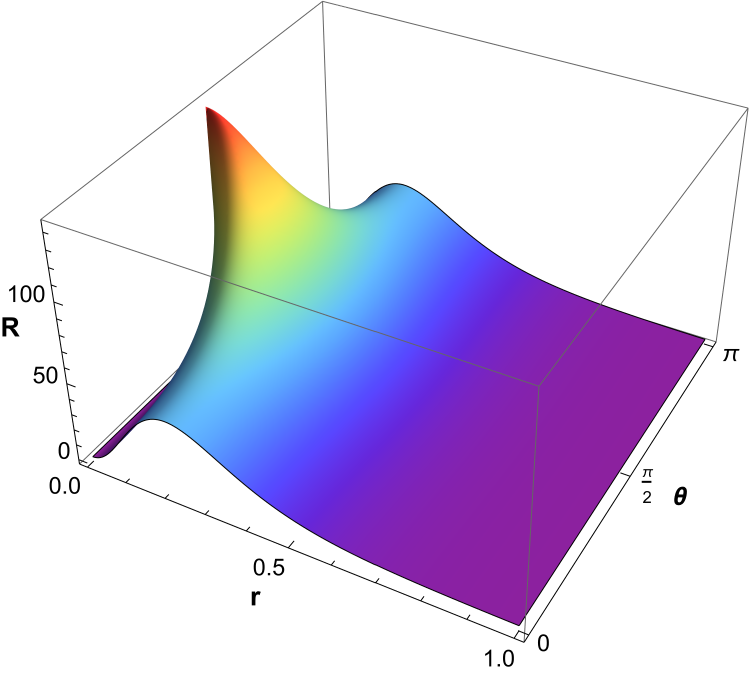}
    \end{subfigure}

    \vspace{0.5em}

    % Second row
    \begin{subfigure}{0.31\textwidth}
        \centering
        \includegraphics[width=\linewidth]{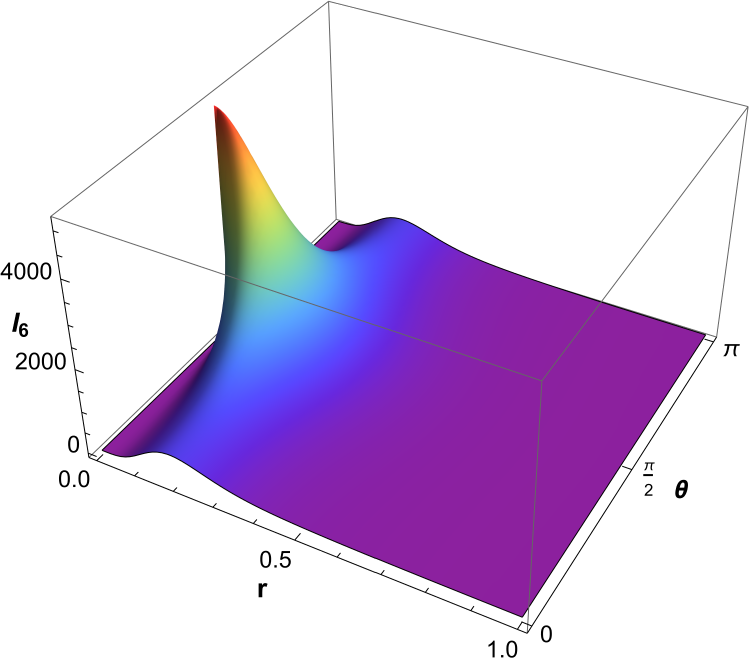}
    \end{subfigure}
    \hfill
    \begin{subfigure}{0.31\textwidth}
        \centering
        \includegraphics[width=\linewidth]{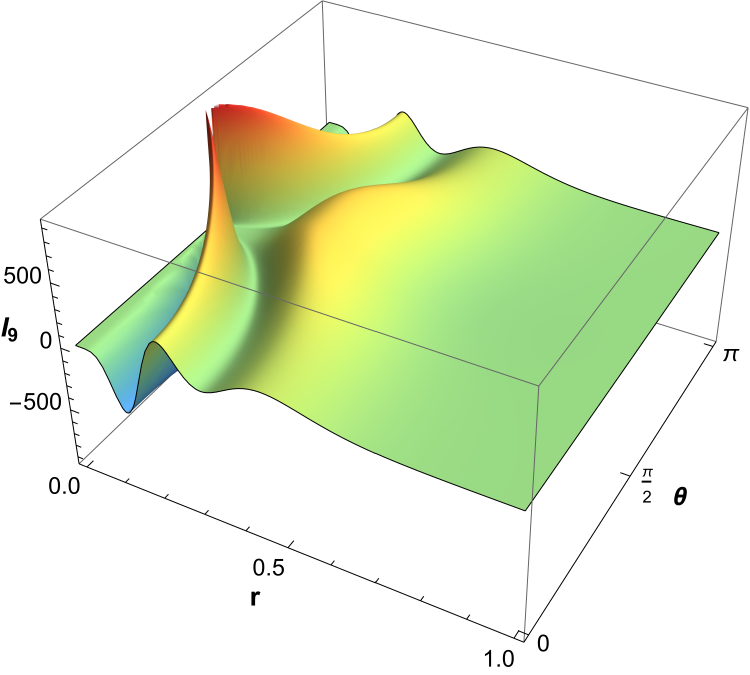}
    \end{subfigure}
    \hfill
    \begin{subfigure}{0.31\textwidth}
        \centering
        \includegraphics[width=\linewidth]{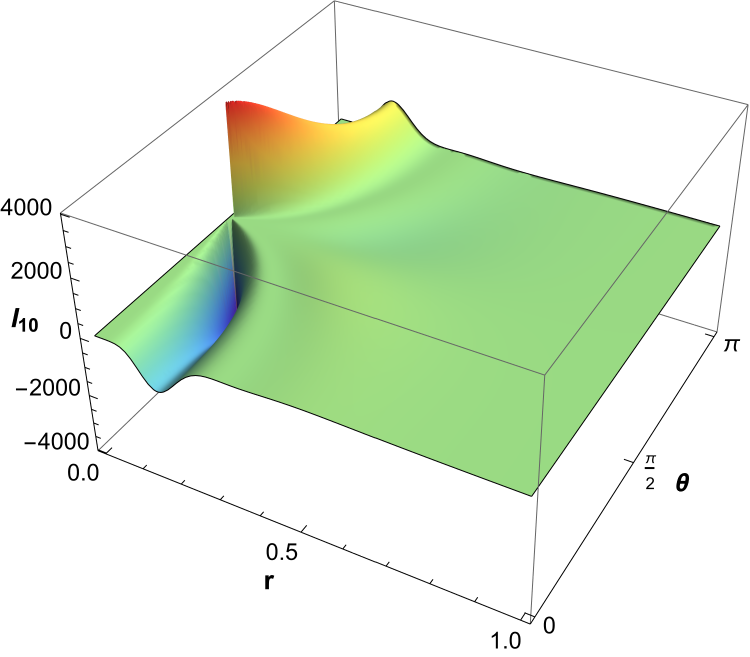}
    \end{subfigure}

    \caption{Scalar invariants  $I_1$, $I_2$, $R$, $I_6$ $I_9$, and $I_{10}$ for rotating ABG metric with $a=0.2$, $q=0.5$ and $M=1$.}
    \label{fig:ZM_inv}
\end{figure}

\noindent Smooth, continuous behaviour as well as finite values of all curvature invariants throughout the domains of all coordinates are a necessity for any non-singular geometry. In general, seventeen Riemann curvature invariants may be required to characterise a four-dimensional spacetime.
However, in \cite{Zakhary1997}, the authors showed that a minimally independent set of invariants for a Petrov type D spacetime with Segre type $[(1,1) (1 1)]$ can be formed by the Weyl invariants $I_1$, $I_2$, the Ricci invariants $I_5(=R)$, $I_6$, and the mixed invariants $I_9$, $I_{10}$ which are given by
\begin{equation}
\begin{split}
    I_1&= C_{\alpha\beta\mu\nu}C^{\alpha\beta\mu\nu},\quad~ I_2= -C^{\alpha\beta\mu\nu}C^{*}_{\alpha\beta\mu\nu}, \quad R=I_5=g^{\mu\nu}R_{\mu\nu} \\ I_6&= R_{\mu\nu}R^{\mu\nu}, \quad
    I_9= C_{\alpha \beta \mu}{}^{\nu} R^{\beta \mu}R_{\nu}{}^{\alpha},
     \quad~ I_{10}= -C^{*}{}_{\alpha \beta \mu}{}^{\nu} R^{\beta \mu}R_{\nu}{}^{\alpha},\\
    \end{split}
    \label{eqs:curvature_invariants}
\end{equation}
where $C_{\alpha\beta\mu\nu}$ and $C^{*}_{\alpha\beta\mu\nu}$ are the Weyl tensor and the dual Weyl tensor defined by
\begin{equation}
\begin{split}
    C_{\alpha\beta\mu\nu}&= R_{\alpha\beta\mu\nu} + \frac{R}{6} \left(g_{\alpha\mu}g_{\beta\nu}-  g_{\alpha\nu}g_{\beta\mu}\right) - \frac{1}{2} \left( g_{\alpha\mu}R_{\beta\nu}-  g_{\alpha\nu}R_{\beta\mu} - g_{\beta\mu}R_{\alpha\nu}+  g_{\beta\nu}R_{\alpha\mu}\right),\\
     C^{*}_{\alpha\beta\mu\nu}&= \frac{1}{2} \sqrt{-g} \epsilon_{\alpha \beta \rho \sigma} C^{\rho\sigma}{}_{\mu\nu}. 
\end{split}    
\end{equation}
 All other Riemann curvature invariants can be expressed in terms of these six invariants. The Kretschmann scalar
 $K=R_{\alpha\beta\mu\nu}R^{\alpha\beta\mu\nu}= I_1+ 2I_6 - R^2/3$. For the rotating ABG spacetime
 \begin{equation*}
     \lim_{r\rightarrow0} \left( \lim_{\theta\rightarrow\pi/2} R \right) =   -\frac{12(q-2M)}{q^3}, \quad     \lim_{r\rightarrow0} \left( \lim_{\theta\rightarrow\pi/2} K \right) =   \frac{24 (q-2M)^2}{q^ 6}, \quad \lim_{r\rightarrow0} \left( \lim_{\theta\rightarrow\pi/2} I_6 \right) =   \frac{36 (q-2M)^2}{q^ 6}.
 \end{equation*}
All other scalar invariants mentioned above become zero in the ($r \rightarrow 0$) limit. For the rotating black hole to be non-singular, all six scalar invariants must be finite in the coordinate space ($r$, $\theta$). In Fig.~\ref{fig:ZM_inv}, we can see that none of the invariants diverges if $q\neq 0$, thus ensuring the non-singular nature of the rotating ABG spacetime. At the ($r \rightarrow 0$, $ \theta \rightarrow \pi/2$) limit, all six scalar invariants corresponding to the rotating BPR black hole become zero. We have also verified that these invariants do not diverge in the ($r$, $\theta$) coordinate space.

% \subsubsection{Energy conditions}

\noindent In Einstein's general relativity, the energy conditions serve as criteria to ensure the physical viability of solutions to Einstein's field equations. In order to check the energy conditions, we consider the energy-momentum tensor $T_{\mu\nu}$. However, as the metric contains an off-diagonal term, one needs to choose the observer in a Locally Non-rotating Frame (LNRF) \cite{Bardeen1972RBH, Bambi2013} where the one-forms of the dual basis of the orthonormal tetrad are
\begin{equation}
\textbf{e}^{(0)}= \left\vert- g_{tt} + \frac{g_{t\phi}^2}{g_{\phi\phi}} \right\vert^{1/2}dt, \quad~ \textbf{e}^{(1)}= \sqrt{g_{rr}}\,dr, \quad~ \textbf{e}^{(2)}=\sqrt{g_{\theta\theta}}d\,\theta, 
\quad~ \textbf{e}^{(3)}= \frac{g_{t\phi}}{\sqrt{g_{\phi\phi}}}\,dt + \sqrt{g_{\phi\phi}}\,d\phi.
\label{NLRF basis}
\end{equation}
\noindent The energy-momentum tensor $T^{\mu\nu}$ in the orthogonal basis is given by
\begin{equation}
    T^{\mu\nu}= \rho\, \hat{e}^{\mu}{}_0\hat{e}^{\nu}{}_0 +p_1\, \hat{e}^{\mu}{}_1\hat{e}^{\nu}{}_1 + p_2\, \hat{e}^{\mu}{}_2\hat{e}^{\nu}{}_2 + p_3\,  \hat{e}^{\mu}{}_3\hat{e}^{\nu}{}_3
\end{equation}
where $\hat{e}^{\mu}{}_a$ are the inverse tetrad components such that
$\textbf{e}^{(a)}=\hat{e}_{\mu}{}^a dx^{\mu}$. Here, the Latin index $a$ denotes the local inertial frame ($a=\lbrace 0, 1, 2, 3\rbrace$).

\noindent Here, $\rho$ is the energy density given by
\begin{equation}
    \rho = T^{\mu\nu} \hat{e}_{\mu}{}^0\hat{e}_{\nu}{}^0,
\end{equation}
while $p_1$, $p_2$ and $p_3$ are the three principal pressures
\begin{equation}
    p_i= T^{\mu\nu} \hat{e}_{\mu}{}^i\hat{e}_{\nu}{}^i, \quad i=1,2,3.
\end{equation}

\begin{figure}[H]
    \centering
    \includegraphics[width=1\linewidth]{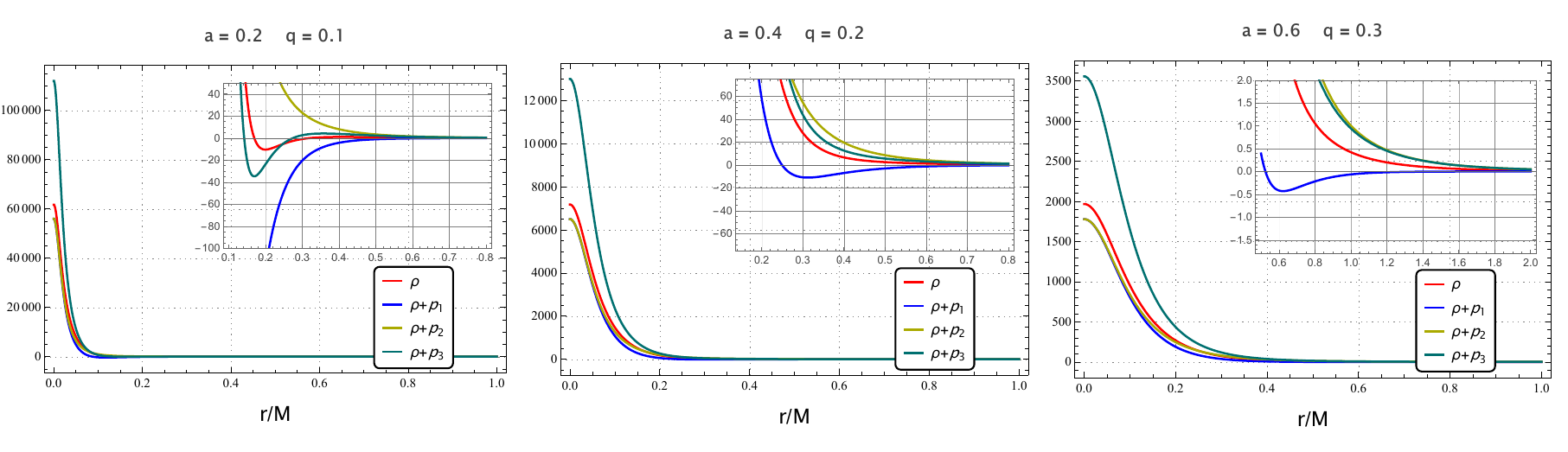}
    \caption{$\rho$ and $\rho+p_i$ for different values of $a$ and $q$ for rotating ABG regular black hole. Here $\theta=\pi/2$ and $M=1$.}
    \label{fig:em_tensorABG}
\end{figure}

\noindent Having computed the energy-momentum tensor, we can now check the weak energy condition (WEC), which requires $\rho \geq 0$ and $\rho+p_i \geq 0$ ($i=1,2,3$). In Fig~\ref{fig:em_tensorABG}, we can observe the variation of $\rho$ and $\rho+p_i$ with the radial coordinate of the rotating ABG geometry for different values of $a$ and $q$ with $\theta=\pi/2$. One can notice that components of the energy-momentum tensor do not diverge, but the WEC is violated, as for a certain range of $r/M$ the values of $\rho$ and $\rho+p_i$ become negative. The WEC is also violated for the rotating BPR regular black hole.

\section{Scalar perturbation and quasinormal modes}\label{QNM}

\noindent QNMs are the characteristic damped oscillations of perturbations in spacetime and can arise due to scalar, fermionic, electromagnetic or gravitational perturbations. In the context of black holes, these QNMs act as a fingerprint of the underlying geometry by providing a direct probe of their mass, angular momentum, and define the approach to stability for the perturbed spacetime. The study of QNMs in black holes gained momentum with the pioneering work of Vishveshwara \cite{Vishveshwara1970zz}, who demonstrated the scattering of gravitational radiation by the Schwarzschild black hole. This was followed by the first systematic calculation of Schwarzschild QNM frequencies by Chandrasekhar and Detweiler \cite{Chandrasekhar1975zza}, while the development of the Teukolsky formalism provided a major breakthrough for perturbation of the Kerr black hole \cite{Teukolsky1973Perturbations}. The subsequent development of numerical and semi-analytical techniques established QNMs as a powerful probe of strong-field gravity, with gravitational wave observations \cite{LIGOScientific2016aoc} opening the possibility of black hole spectroscopy. In the context of this work, we extend the study of QNMs to the scalar perturbations of charged rotating regular black holes. The equation of motion of a massless scalar perturbation field $\Phi$ is governed by the Klein-Gordon equation, which can be written as
\begin{equation}
    \Box \Phi= \frac{1}{\sqrt{-g}}\partial_\mu(\sqrt{-g}g^{\mu\nu}\partial_{\nu}\Phi)=0. \label{kg}
\end{equation}
Assuming separation of variables and putting $\Phi(t,r,\theta,\phi)=e^{-i\omega t}e^{i m \phi}S(\theta)R(r)$ into Eq.~\eqref{kg} we get
\begin{align}
    &\left(\Delta(r)R''(r)+\Delta'(r)R'(r)\right)S(\theta)+R(r)(S''(\theta)+cot(\theta)S'(\theta)) \nonumber\\ 
    &+\frac{1}{2} \left(a^2 \omega ^2 \cos (2 \theta )+\frac{2 a^2 m^2+\omega ^2 \left(a^2+r^2\right) \left(a^2+2 r^2\right)+2 a r \omega  M(r) (a \omega -4 m)}{a^2-2 r M(r)+r^2}-2 m^2 \csc ^2(\theta )\right)R(r)S(\theta)=0.
\end{align}
This equation can be separated into radial and angular equations
\begin{equation}
    \Delta(r)R''(r)+\Delta'(r)R'(r)
    +\left(\frac{ (a^2+r^2)^2\omega^2+a^2 m^2-4 a m \omega r M(r)}{\Delta(r)}-a^2 \omega ^2-A_{lm}\right)R(r)=0, \label{Radial}
\end{equation}
\begin{equation}
    S''(\theta)+cot(\theta)S'(\theta)+\left(a^2 \omega ^2 \cos ^2(\theta )-m^2 \csc ^2(\theta )+A_{lm}\right)S(\theta)=0,
\end{equation}
where $A_{lm}$ is the separation constant. Now that we have the perturbation equations, we can compute the QNMs of the spacetime. We will employ the spectral and WKB methods to compute the QNMs for these geometries. There are two equations and two unknowns, namely $\omega$ and $A_{lm}$. Therefore, we will solve this problem iteratively. First, we will solve the radial equation for $\omega$ assuming $A_{lm}=l(l+1)$, where $l$ is the multipole number. Next, we will solve the angular equation for the separation constant $A_{lm}$. Next, we update the radial equation with this new value of $A_{lm}$ and get a new $\omega$, which is again utilised in the angular equation to get an updated value of $A_{lm}$ for the subsequent iteration. Thus, we iteratively update both $\omega$ and $A_{lm}$ until the corrections from further iterations are negligible. This iterative procedure is followed for both methods. However, we use the WKB method only for the radial equation.

\subsection{Quasinormal modes}

\noindent Having obtained the scalar perturbation equations for the charged rotating regular black holes, we can now compute the QNM spectra using the two methods detailed in Appendix \ref{methods}. In Tables \ref{tab:qnm_0} and \ref{tab:qnm_mm}, we present the fundamental modes of rotating ABG and rotating BPR regular black holes corresponding to $l=m=2$ and $l=-m=2$ cases, respectively. The results obtained from the spectral method and sixth-order WKB are in good agreement. However, setting $q=0$ and comparing with the scalar QNM frequencies of the Kerr black hole \cite{Berti2009kk, Berti2005ys, berti2025ringdown}, we find that the spectral method is more accurate. Therefore, in all subsequent plots and tables, we will use the QNM frequencies obtained through the spectral method.

\begin{table}[H]
    \centering
    \footnotesize
    \setlength{\tabcolsep}{4pt}
    \renewcommand{\arraystretch}{1.2}
    \singlespacing
    \scalebox{1}{
    \begin{tabular}{|c|c||c|c||c|c|}
        \hline
         & & \multicolumn{2}{c||}{\textbf{rotating ABG}} & \multicolumn{2}{c|}{\textbf{rotating BPR}}\\
        \hline
        $q$ & $a$ & Spectral method & WKB 6 & Spectral method & WKB 6  \\
        \hline
        0.1 & 0.1 & $0.501328-0.0965811i$ & $0.501325 - 0.0965896i$  & $0.500409-0.0967099i$ & $0.500406 - 0.0967178i$  \\
            & 0.3 & $0.539389-0.0956722i$ & $0.539385 - 0.0956848i$  & $0.538144-0.0958629i$ & $0.538141 - 0.0958748i$  \\
            & 0.5 & $0.589393-0.0931468i$ & $0.589388 - 0.0931663i$  & $0.587554-0.0934708i$ & $0.587549 - 0.0934896i$  \\
            & 0.7 & $0.661694-0.0867914i$ & $0.661692 - 0.0868151i$  & $0.65848-0.0874857i$  & $0.658479 - 0.0875083i$  \\
            & 0.9 & $0.796415-0.0654719i$ & $0.796471 - 0.0654016i$  & $0.786959-0.0683539i$ & $0.787012 - 0.068298i$   \\
        \hline
        0.3 & 0.1 & $0.517333-0.0956273i$ & $0.517329 - 0.0956399i$  & $0.508635-0.0969763i$ & $0.508633 - 0.0969844i$  \\
            & 0.3 & $0.560744-0.0938444i$ & $0.560739 - 0.0938612i$  & $0.548608-0.0959202i$ & $0.548605 - 0.0959323i$  \\
            & 0.5 & $0.620839-0.0892014i$ & $0.620832 - 0.089226i$   & $0.601868-0.0930162i$ & $0.601863 - 0.0930345i$  \\
            & 0.7 & $0.719724-0.0752931i$ & $0.719729 - 0.0753242i$  & $0.681113-0.0854719i$ & $0.681114 - 0.0854929i$  \\
        \hline
    \end{tabular}
    }
    \caption{QNM fundamental mode frequencies computed using the spectral method and sixth-order WKB for rotating ABG and rotating BPR black holes. Here $l=2$, $m=2$ and $M=1$.}
    \label{tab:qnm_0}
\end{table}

\begin{table}[H]
    \centering
    \footnotesize
    \setlength{\tabcolsep}{4pt}
    \renewcommand{\arraystretch}{1.2}
    \singlespacing
    \scalebox{1}{
    \begin{tabular}{|c|c||c|c||c|c|}
        \hline
         & & \multicolumn{2}{c||}{\textbf{rotating ABG}} & \multicolumn{2}{c|}{\textbf{rotating BPR}}\\
        \hline
        $q$ & $a$ & Spectral method & WKB 6 & Spectral method & WKB 6  \\
        \hline
        0.1 & 0.1 & $0.470779-0.0966633i$ & $0.470777 - 0.0966712i$  & $0.470062-0.0967595i$ & $0.470061 - 0.096767i$   \\
            & 0.3 & $0.445402-0.0962757i$ & $0.445401 - 0.0962845i$  & $0.444823-0.0963535i$ & $0.444822 - 0.096362i$   \\
            & 0.5 & $0.423802-0.0956028i$ & $0.423801 - 0.0956124i$  & $0.423321-0.0956698i$ & $0.423321 - 0.0956793i$  \\
            & 0.7 & $0.405074-0.0947501i$ & $0.405074 - 0.0947607i$  & $0.404667-0.0948105i$ & $0.404667 - 0.0948211i$  \\
            & 0.9 & $0.388599-0.0937834i$ & $0.388599 - 0.0937954i$  & $0.388251-0.0938394i$ & $0.388251 - 0.0938513i$  \\
        \hline
        0.3 & 0.1 & $0.483479-0.0961288i$ & $0.483477 - 0.0961401i$  & $0.476824-0.0971147i$ & $0.476822 - 0.0971226i$  \\
            & 0.3 & $0.455864-0.0959544i$ & $0.455863 - 0.0959654i$  & $0.450552-0.0967411i$ & $0.450551 - 0.0967501i$  \\
            & 0.5 & $0.432649-0.0953902i$ & $0.432648 - 0.0954012i$  & $0.428283-0.0960608i$ & $0.428283 - 0.0960706i$  \\
            & 0.7 & $0.412699-0.094591i$  & $0.412699 - 0.0946028i$  & $0.409037-0.0951894i$ & $0.409037 - 0.0952001i$  \\
        \hline
    \end{tabular}}
    \caption{QNM fundamental mode frequencies computed using the spectral method and sixth-order WKB for rotating ABG and rotating BPR black holes. Here $l=2$, $m=-2$ and $M=1$.}
    \label{tab:qnm_mm}
\end{table}

\begin{figure}[H]
    \centering

    \begin{subfigure}{0.48\textwidth}
        \centering
        \includegraphics[width=\linewidth]{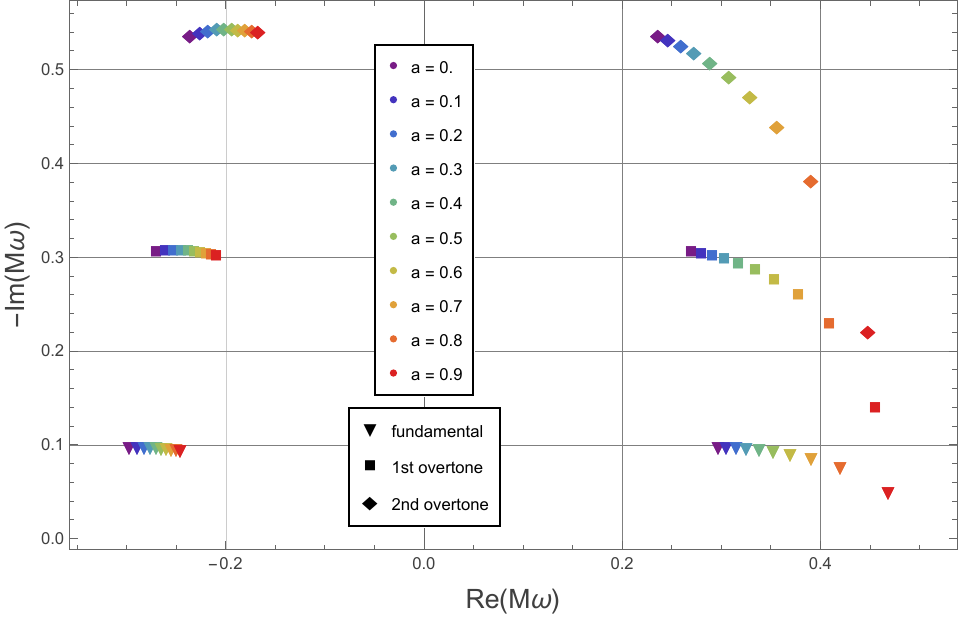}
    \end{subfigure}
    \hfill
    \begin{subfigure}{0.48\textwidth}
        \centering
        \includegraphics[width=\linewidth]{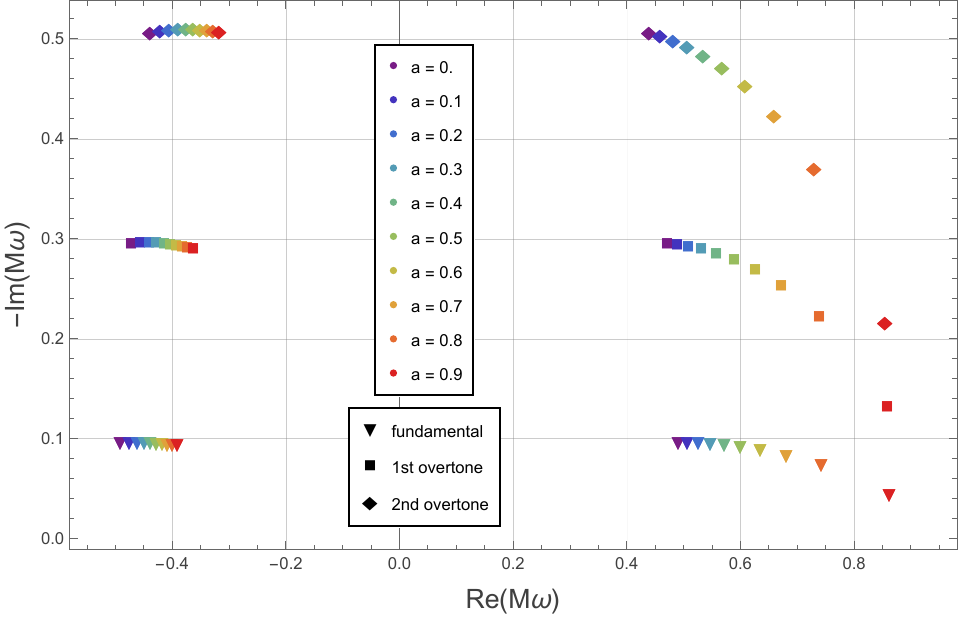}
    \end{subfigure}

    \caption{QNM spectra of rotating ABG black hole for $q=0.2M$. (left) For $l=1,m=1$ case. (right) For $l=2,m=2$ case.}
    \label{fig:ABG}
\end{figure}

\begin{figure}[H]
    \centering

    \begin{subfigure}{0.48\textwidth}
        \centering
        \includegraphics[width=\linewidth]{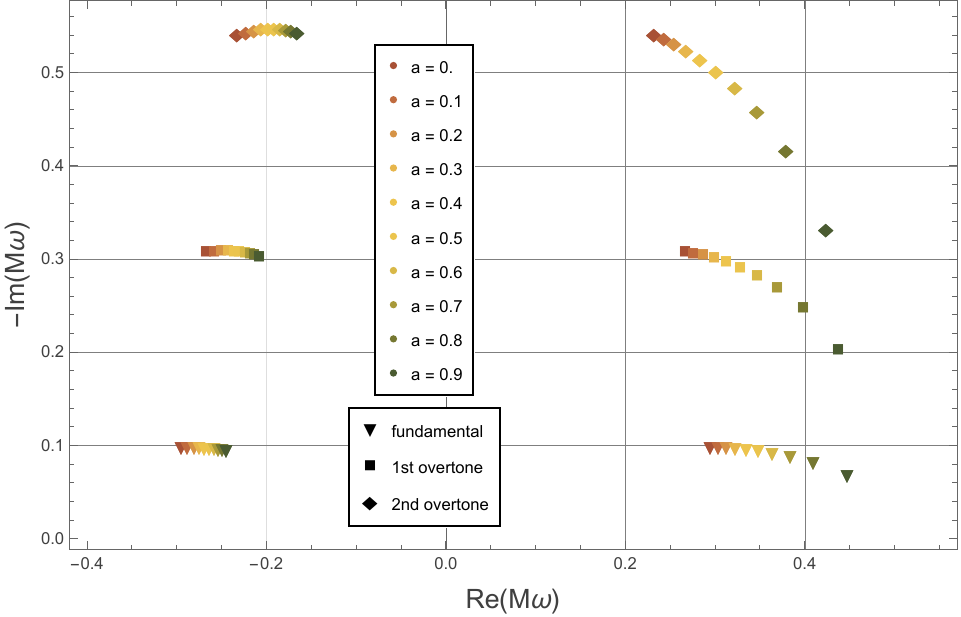}
    \end{subfigure}
    \hfill
    \begin{subfigure}{0.48\textwidth}
        \centering
        \includegraphics[width=\linewidth]{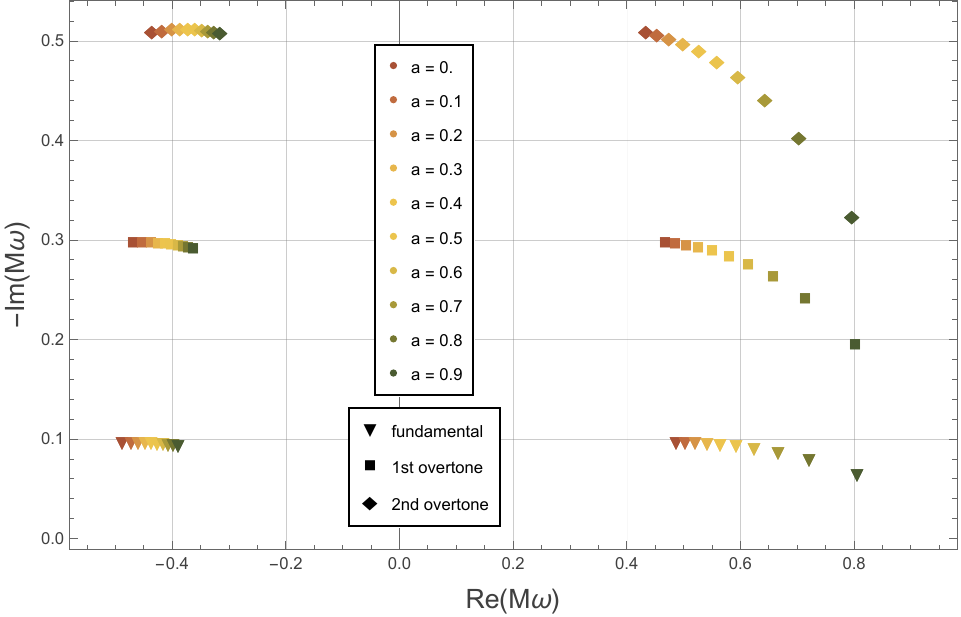}
    \end{subfigure}

    \caption{QNM spectra of rotating BPR black hole for $q=0.2M$. (left) For $l=1,m=1$ case. (right) For $l=2,m=2$ case.}
    \label{fig:BPR}
\end{figure}

\noindent In Fig.~\ref{fig:ABG}, we have plotted the QNMs corresponding to the scalar perturbation of the rotating ABG spacetime (with $q=0.2M$) for the $l=1$, $m=1$ and $l=2$, $m=2$ case. Here, one can observe that although the actual values of the QNM frequencies differ, the qualitative behaviour in both cases is very similar. The QNM spectra for the Rotating BPR black hole with $q=0.2M$ are plotted in Fig.~\ref{fig:BPR}. We will compare these spectra with that of the Kerr--Newman black hole \cite{Berti2005eb, Dias2015wqa, Dias2021yju}. \\

\noindent Fig.~\ref{fig:ABG} and \ref{fig:BPR} illustrate the evolution of the QNM spectrum in the complex-frequency plane as the rotation parameter $a$ is varied from 0 to 0.9. Similar to other rotating geometries, here we observe that a distinct separation between modes with positive and negative values of Re($\omega$) emerges for non-zero values of the rotation parameter $a$. A more pronounced rotation dependence is observed in the real part of the QNM frequencies, with the positive-frequency branch shifting toward larger $|$Re($\omega$)$|$ as the rotation increases. However, for the branch with a negative real part, $|$Re($\omega$)$|$ reduces with increasing values of $a$. Thus, the rotation of the black hole introduces a rotational splitting, and the QNM spectra lose their mirror symmetry (Absolute values of real parts are no longer the same for the two branches). The fundamental mode exhibits only a comparatively weak variation in its damping rate. In contrast, the first and second overtones display a stronger dependence on $a$, with their damping rates decreasing significantly as the rotation increases. This indicates that the higher overtones are more sensitive to the rotational properties of the spacetime than the fundamental mode.\\

\begin{figure}[H]
    \centering

    \begin{subfigure}{0.45\textwidth}
        \centering
        \includegraphics[width=\linewidth]{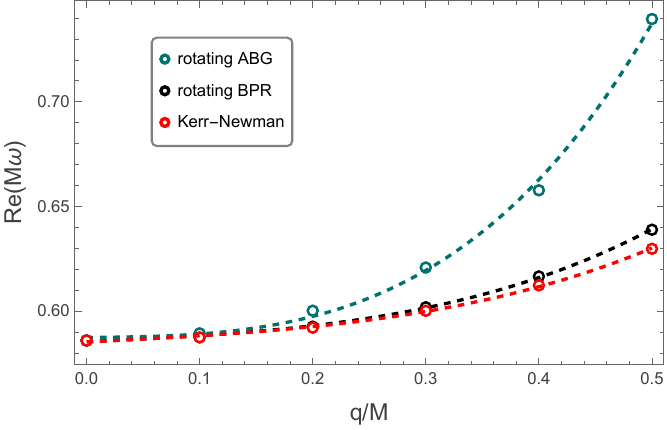}
    \end{subfigure}
    \qquad
    \begin{subfigure}{0.45\textwidth}
        \centering
        \includegraphics[width=\linewidth]{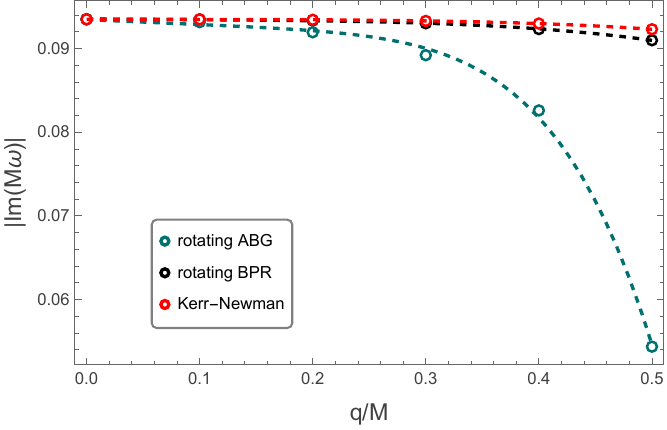}
    \end{subfigure}

    \caption{Dependence of fundamental mode on the charge parameter $q$. (left) For the real part. (right) For the absolute value of the imaginary part. The points are the fundamental modes corresponding to $a=0.5M$, $l=2,m=2$; the dashed lines are analytical fits.}
    \label{fig:q_dep}
\end{figure}

\begin{figure}[H]
    \centering

    \begin{subfigure}{0.45\textwidth}
        \centering
        \includegraphics[width=\linewidth]{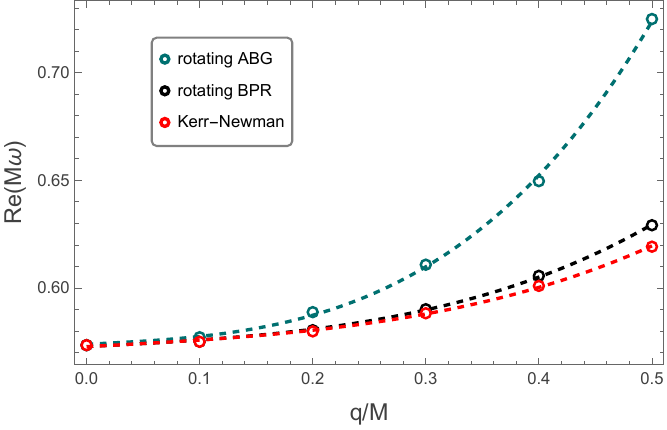}
    \end{subfigure}
    \qquad
    \begin{subfigure}{0.45\textwidth}
        \centering
        \includegraphics[width=\linewidth]{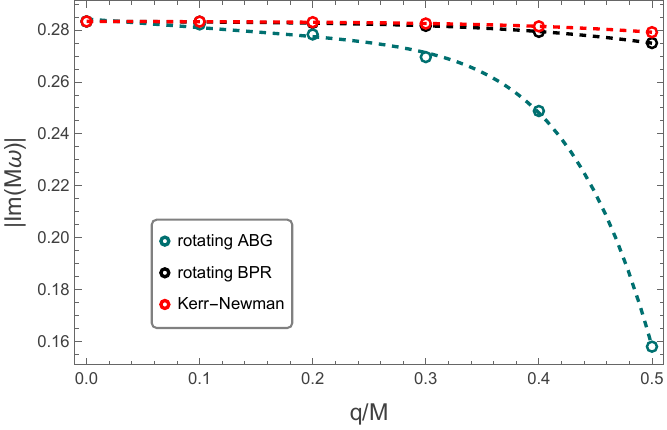}
    \end{subfigure}

    \caption{Dependence of first overtone on the charge parameter $q$. (left) For the real part. (right) For the absolute value of the imaginary part. The points are the first overtones corresponding to $a=0.5M$, $l=2,m=2$; the dashed lines are analytical fits.}
    \label{fig:q_dep1}
\end{figure}

\begin{figure}[H]
    \centering

    \begin{subfigure}{0.45\textwidth}
        \centering
        \includegraphics[width=\linewidth]{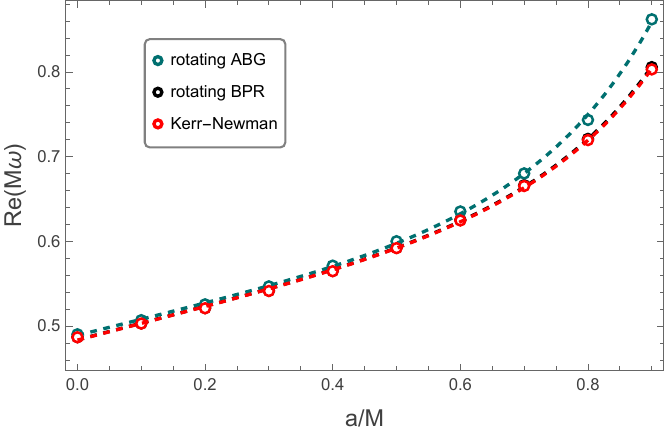}
    \end{subfigure}
    \qquad
    \begin{subfigure}{0.45\textwidth}
        \centering
        \includegraphics[width=\linewidth]{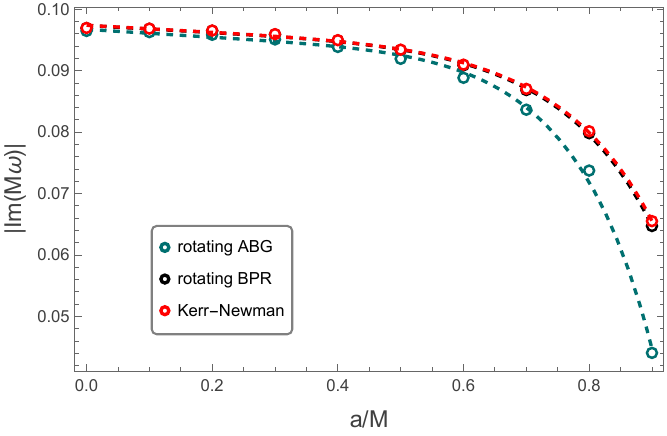}
    \end{subfigure}

    \caption{Dependence of fundamental mode on the rotation parameter $a$. (left) For the real part. (right) For the absolute value of the imaginary part. The points are the fundamental modes corresponding to $q=0.2M$, $l=2$, and $m=2$; the dashed lines are analytical fits.}
    \label{fig:a_dep}
\end{figure}

\begin{figure}[H]
    \centering

    \begin{subfigure}{0.45\textwidth}
        \centering
        \includegraphics[width=\linewidth]{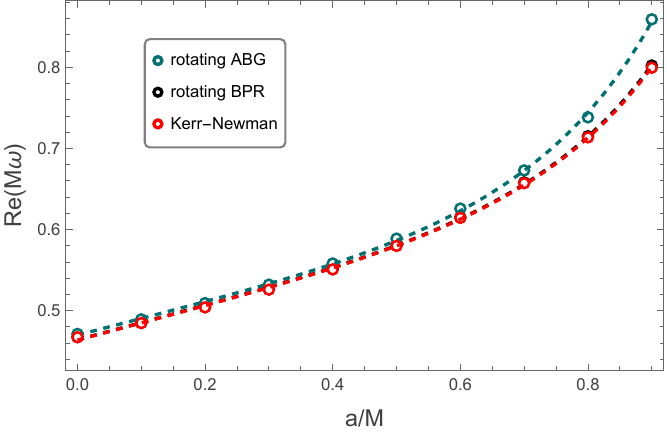}
    \end{subfigure}
    \qquad
    \begin{subfigure}{0.45\textwidth}
        \centering
        \includegraphics[width=\linewidth]{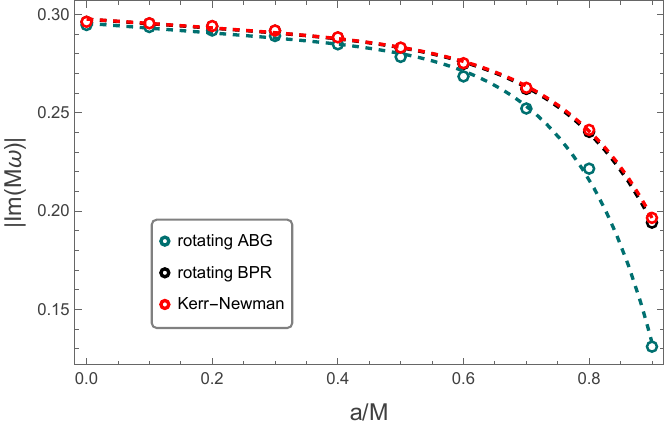}
    \end{subfigure}

    \caption{Dependence of first overtone on the rotation parameter $a$. (left) For the real part. (right) For the absolute value of the imaginary part. The points are the first overtones corresponding to $q=0.2M$, $l=2$, and $m=2$; the dashed lines are analytical fits.}
    \label{fig:a_dep1}
\end{figure}

\noindent For both the rotating charged regular black holes, the real parts of the QNMs (positive-frequency branch) increase with both the rotational parameter $a$ and the charge parameter $q$, whereas the increase of $a$ and $q$ results in a decrease of the absolute value of the imaginary parts, similar to the Kerr--Newman black hole. We illustrate these features more clearly by plotting the variations of Re($\omega$) and $|$Im($\omega$)$|$ as functions of the charge parameter $q$ (see Fig.~\ref{fig:q_dep},  \ref{fig:q_dep1})  and the rotational parameter $a$ (see Fig.~\ref{fig:a_dep},  \ref{fig:a_dep1}) separately.

 \noindent Fig.~\ref{fig:q_dep} and Fig.~\ref{fig:q_dep1} show that in all three cases (rotating ABG, rotating BPR, and Kerr--Newman), Re($\omega$) increases steadily while the absolute value of Im($\omega$) decreases as we increase the charge $q$. The pattern holds for both fundamental modes and first overtones. The three curves are very close at low values of $q/M$, indicating that the spectra differ only slightly when the charge parameter is small. As $q$ increases, the separation becomes more noticeable, with the rotating ABG case giving the highest values of Re($\omega_{QNM}$), followed by rotating BPR, while Kerr--Newman remains slightly lower. The values of $|$Im($\omega$)$|$ are lower for the rotating ABG spacetime than the other two spacetimes for the same value of $q/M$. In the $a=0.5M$, $q=0.5M$ case, the real and imaginary parts of QNM fundamental mode corresponding to the rotating BPR black hole differ from those of the Kerr--Newman black hole by $1.45\%$ and $1.42\%$ respectively. For the rotating ABG spacetime, these relative changes are $17.43\%$ and $41.09\%$, respectively. \\

 \noindent With increasing spin ($a$) of the black holes, the real part of QNM increases while the absolute value of the imaginary part decreases for both fundamental modes (Fig.~\ref{fig:a_dep}) and first overtones (Fig.~\ref{fig:a_dep1}). Here, one observes that the QNMs of the rotating BPR and Kerr--Newman black holes are almost indistinguishable. However, for higher values of $a/M$, the QNM spectra of the rotating ABG metric deviate slightly. For the fundamental modes, in the $a=0.9M$, $q=0.2M$ case, the rotating BPR black hole exhibits relative deviation of $0.33\%$ and $1.21\%$ in the real and imaginary parts, respectively, compared to the Kerr--Newman black hole. For the rotating ABG spacetime, the corresponding deviations increase to $7.36\%$ and $32.74\%$, respectively.

\noindent So far, we have shown the dependence of the QNM spectra for the $l=m=2$ case on the charge and spin parameter of the three spacetimes. However, these dependencies change for negative values of the azimuthal number $m$.

\begin{figure}[H]
    \centering

    \begin{subfigure}{0.45\textwidth}
        \centering
        \includegraphics[width=\linewidth]{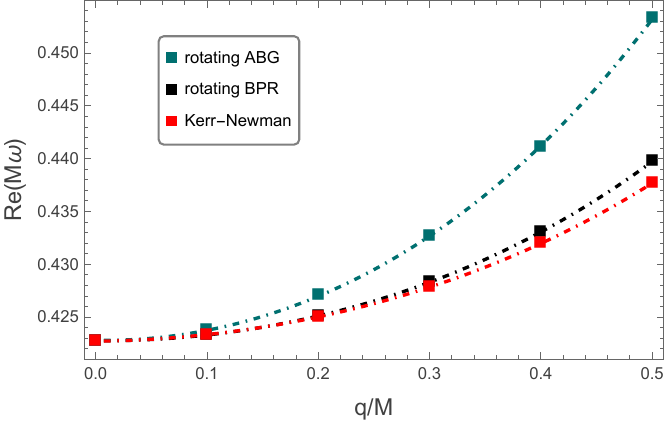}
    \end{subfigure}
    \qquad
    \begin{subfigure}{0.45\textwidth}
        \centering
        \includegraphics[width=\linewidth]{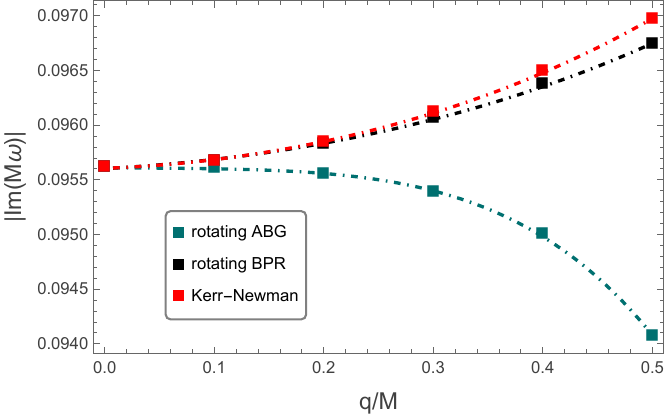}
    \end{subfigure}

    \caption{Dependence of fundamental mode ($l=2$, $m=-2$) on the charge parameter $q$. (left) For the real part. (right) For the absolute value of the imaginary part. The points are the fundamental modes corresponding to $a=0.5M$, $l=2$, and $m=-2$; the dot-dashed lines are analytical fits.}
    \label{fig:q_depm}
\end{figure}

\noindent  A clear and distinct dependence of fundamental mode ($l=2$, $m=-2$) on charge parameter $q$ is shown in Fig.~\ref{fig:q_depm}. The real part increases with $q$ for all three geometries. This growth is most pronounced in the rotating ABG spacetime, while the rotating BPR and Kerr--Newman cases exhibit comparatively weaker variations. By contrast, the absolute values of the imaginary parts show qualitatively different behaviour. It decreases as the value of the charge parameter increases in the rotating ABG black hole, whereas it increases in both the rotating BPR and Kerr--Newman spacetimes. Thus, the charge parameter dependence of the damping rate in the negative azimuthal number ($m$) case is sensitive to the underlying regularisation of the black hole geometry. The separation of the ABG mode from the BPR and Kerr--Newman results at larger $q/M$ highlights the distinctive effect of the regular core on the quasinormal spectrum.

\begin{figure}[H]
    \centering

    \begin{subfigure}{0.45\textwidth}
        \centering
        \includegraphics[width=\linewidth]{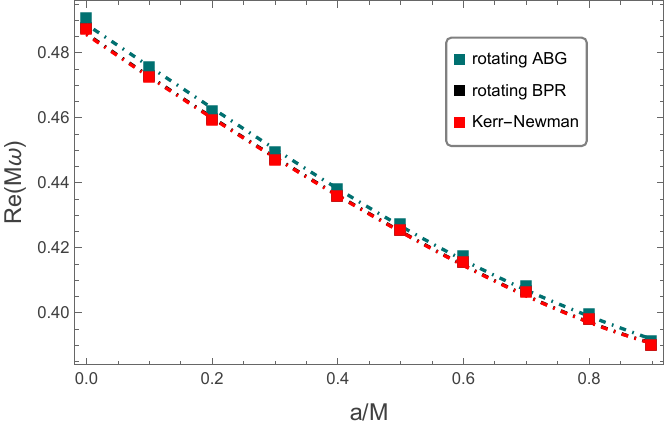}
    \end{subfigure}
    \qquad
    \begin{subfigure}{0.45\textwidth}
        \centering
        \includegraphics[width=\linewidth]{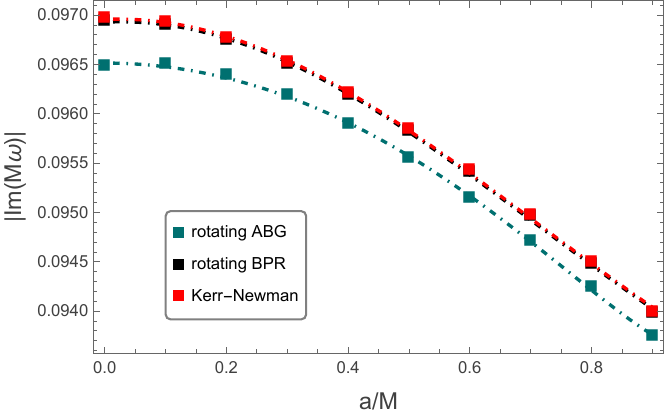}
    \end{subfigure}

    \caption{Dependence of fundamental mode  ($l=2$, $m=-2$) on the rotation parameter $a$. (left) For the real part. (right) For the absolute value of the imaginary part. The points are the fundamental modes corresponding to $q=0.2M$, $l=2$, and $m=-2$; the dot-dashed lines are analytical fits.}
    \label{fig:a_depm}
\end{figure}

\noindent From Fig.~\ref{fig:a_depm}, we observe that no such qualitative distinction between the geometries is present for the spin dependence. As we increase the value of the spin parameter $a$, $|$Im($\omega$)$|$ decreases for all three spacetimes, similar to what we observed for the $l=m=2$ case. However, unlike the $l=m=2$ case, the value of Re($\omega$) in the $l=2$ and $m=-2$ case decreases with increasing spin.\\

\noindent Here we summarise the spin and charge dependence of the QNM spectra. For the $l=m=2$ case, all three spacetimes (rotating ABG, rotating BPR, and Kerr--Newman)
show the same qualitative trend: the real part of QNM frequencies increases with both the charge parameter $q$ and spin parameter $a$, while the magnitude of the imaginary part decreases with both parameters. These trends change for the negative azimuthal number case ($l=2$, $m=-2$): though the real part still increases with $q$ for all three geometries, the imaginary part behaves qualitatively differently -- it decreases with $q$ for the rotating ABG black hole but increases for the rotating BPR/Kerr--Newman Black hole. For the $m=-2$ case, both the real part and the absolute value of the imaginary part reduce with increasing spin.

\section{Grey-body factors and superradiance} \label{Grey}

\noindent While studying the spectral properties of a black hole geometry, the grey-body factors (GBFs) are often studied along with the QNMs. While computing the QNMs, we need to impose conditions that ensure purely ``outgoing" waves at the event horizon and spatial infinity, whereas the GBFs correspond to a different boundary condition that permits incoming waves from the event horizon. Thus, the GBFs characterise the scattering properties of gravitational waves by the spacetime. GBFs are an important aspect to study, as they can help estimate remnant parameters from gravitational wave ringdown \cite{Oshita2023cjz}. In this section, we will focus on the scattering of scalar waves in charged rotating regular black hole geometries. To address this scattering problem, we rewrite the radial perturbation equation (Eq.~\eqref{rad_wkb}) in the Schrodinger-like form
\begin{equation}
    \frac{d^2 \Psi}{d r_*^2} + (\Omega^2-V(r_*)) \Psi = 0.
\end{equation}
Here, we use the real frequency of the scattering problem $\Omega$ instead of complex frequencies ($\omega$) associated with QNMs. Due to the symmetry of the scattering problem, the boundary conditions can be expressed (in an equivalent form) as
\begin{equation}
    \Psi \sim
    \begin{cases}
    e^{-i\Omega r_*}+\mathcal{R}e^{+i\Omega r_*},
    & r_* \rightarrow +\infty, \\[6pt]
    \mathcal{T}e^{-i(\Omega-m\Omega_H)r_*},
    & r_* \rightarrow -\infty,
    \end{cases} \label{assym}
\end{equation}
where $\Omega_H$ is the angular velocity at the horizon and $\mathcal{R}$, $\mathcal{T}$ are reflection and transmission coefficients respectively. Since the effective potentials $V(r_*)$ for both the rotating regular geometries are single-peaked, we can employ the WKB method. The GBFs can be computed using \cite{Iyer1987}
\begin{equation}
    \Gamma_{lm}=|\mathcal{T}|^2=\frac{1}{1+e^{2\pi i \mathcal{K}}}, \label{GBF}
\end{equation}
where $\mathcal{K}$ is obtained using the WKB formula
\begin{equation}
    -i\mathcal{K}=\frac{\Omega^2-V_0}{\sqrt{-2V_0''}}-\sum_{j=2}^{k}\Lambda_j(\mathcal{K}). \label{GBF_org}
\end{equation}
Here, $V_0$ and $V_0''$ denote the effective potential and its second derivative with respect to $r_*$ at the peak of the potential ($r_*^{(0)}$), while the $\Lambda_j$ represent correction terms beyond eikonal approximation \cite{Iyer1987, Konoplya2003, Matyjasek2017}. We will use the sixth-order WKB ($k=6)$ to compute the GBFs.\\

\noindent Since the GBFs and QNMs are governed by the same perturbation equations, a natural correspondence arises between them. Konoplya and Zhidenko \cite{Konoplya2024} first established this connection for static, spherically symmetric black holes. The same prescription also applies to the Kerr black hole, as shown in \cite{Huang2025}. Here, we will show that an approximate correspondence between GBFs and QNMs also holds for rotating charged regular spacetimes. As the WKB formulation remains unchanged, we end up with the same relation between GBFs and QNMs as \cite{Konoplya2024}; 
\begin{equation}
    -i\mathcal{K} = -\frac{\Omega^2-\operatorname{Re}(\omega_0)^2}
    {4\operatorname{Re}(\omega_0)\operatorname{Im}(\omega_0)}
    +\Delta_1+\Delta_2+\Delta_f
    +\mathcal{O}\left(l^{-3}\right), \label{GBF_cor}
\end{equation}
where $l$ is the multipole number and $\Delta_1$, $\Delta_2$, $\Delta_f$ are higher order corrections to the eikonal formula (up to sixth order) that encompass all the corrections up to $\mathcal{O}(l^{-2})$. They have the following form \cite{Konoplya2024}
\begin{equation}
    \Delta_1=\frac{\operatorname{Re}(\omega_0)-\operatorname{Re}(\omega_1)}
    {16\operatorname{Im}(\omega_0)}
    +\mathcal{O}\left(l^{-2}\right).
\end{equation}
\begin{equation}
\begin{aligned}
\Delta_2
={}&
-\frac{\Omega^2-\operatorname{Re}(\omega_0)^2}
{32\operatorname{Re}(\omega_0)\operatorname{Im}(\omega_0)}
\left[
\frac{\left(\operatorname{Re}(\omega_0)
-\operatorname{Re}(\omega_1)\right)^2}
{4\operatorname{Im}(\omega_0)^2}
-
\frac{3\operatorname{Im}(\omega_0)-\operatorname{Im}(\omega_1)}
{3\operatorname{Im}(\omega_0)}
\right]
\\[4pt]
&+
\frac{\left(\Omega^2-\operatorname{Re}(\omega_0)^2\right)^2}
{16\operatorname{Re}(\omega_0)^3\operatorname{Im}(\omega_0)}
\left[
1+
\frac{\operatorname{Re}(\omega_0)
\left(\operatorname{Re}(\omega_0)-\operatorname{Re}(\omega_1)\right)}
{4\operatorname{Im}(\omega_0)^2}
\right]
+\mathcal{O}\left(l^{-3}\right).
\end{aligned}
\end{equation}

\begin{equation}
\begin{aligned}
\Delta_f
={}&
\frac{\left(-\Omega^2+\operatorname{Re}(\omega_0)^2\right)^3}
{32\operatorname{Re}(\omega_0)^5\operatorname{Im}(\omega_0)}
\Bigg[
1+
\frac{\operatorname{Re}(\omega_0)
\left(\operatorname{Re}(\omega_0)-\operatorname{Re}(\omega_1)\right)}
{4\operatorname{Im}(\omega_0)^2}
\\[4pt]
&
+\operatorname{Re}(\omega_0)^2
\left(
\frac{\left(\operatorname{Re}(\omega_0)
-\operatorname{Re}(\omega_1)\right)^2}
{16\operatorname{Im}(\omega_0)^4}
-
\frac{3\operatorname{Im}(\omega_0)-\operatorname{Im}(\omega_1)}
{12\operatorname{Im}(\omega_0)}
\right)
\Bigg]
+\mathcal{O}\left(l^{-3}\right).
\end{aligned}
\end{equation}
Here $\omega_0$ and $\omega_1$ are the QNM fundamental mode and first overtone, respectively. Now, we can put the exact values of the QNM frequencies obtained earlier (Sec. \ref{QNM}) in Eq.~\eqref{GBF_cor} to obtain $\mathcal{K}$ as a function of $\Omega$. Independently, we can get $\mathcal{K}$ using Eq.~\eqref{GBF_org}. Finally, we plug in $\mathcal{K}$ obtained from two methods into Eq.~\eqref{GBF} to establish QNM-GBF correspondence.\\

\begin{figure}[H]
    \centering

    \begin{subfigure}{0.48\textwidth}
        \centering
        \includegraphics[width=\linewidth]{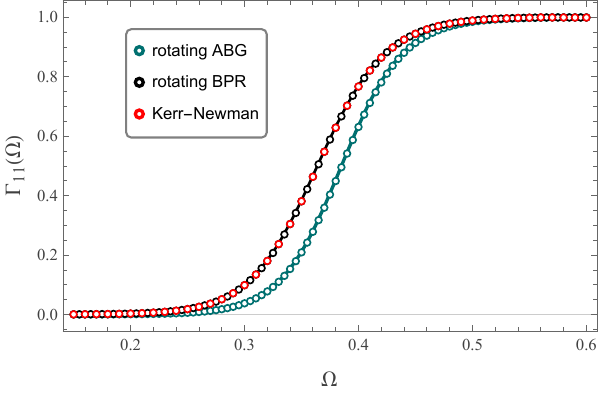}
    \end{subfigure}
    \hfill
    \begin{subfigure}{0.48\textwidth}
        \centering
        \includegraphics[width=\linewidth]{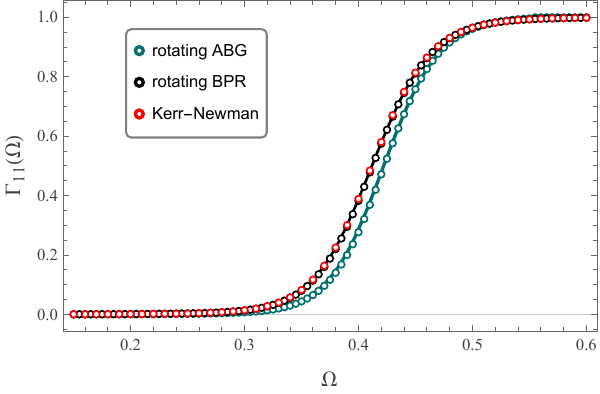}
    \end{subfigure}

    \caption{QNM and GBF correspondence for $l=1$, $m=1$ and $M=1$. (left) $a=0.5$, $q=0.4$, (right) $a=0.8$, $q=0.2$. The points represent the GBFs computed from the perturbation equation (solving Eq.~\eqref{GBF_org}) while the solid lines are plotted using QNM-GBF correspondence (solving Eq.~\eqref{GBF_cor}).}
    \label{fig:GBF}
\end{figure}

\noindent Fig.~\ref{fig:GBF} compares the GBFs ($\Gamma_{11}$) as a function of the frequency $\Omega$ for the rotating ABG, rotating BPR, and Kerr--Newman black holes. All three geometries show a monotonic increase in the transmission probability from nearly zero at low frequencies to unity at high frequencies. The rotating BPR and Kerr--Newman \cite{Boonserm2014rma} black holes exhibit remarkably similar GBFs, with their curves almost overlapping throughout the $\Omega$ range. However, the GBFs for rotating ABG spacetime are suppressed at intermediate frequencies, and their transition toward unity is shifted to higher frequencies. This indicates that the effective potential experienced by the perturbation in the rotating ABG spacetime provides a stronger barrier compared with the other two geometries. These differences diminish at sufficiently high frequencies, where all three GBFs asymptotically approach unity. Thus, the GBF provides a clear distinction between the rotating ABG geometry and the Kerr--Newman/rotating BPR black holes, particularly in the intermediate frequency regime.

\subsection{Superradiance}

\noindent A rotating regular black hole can act as an amplifier: a scalar wave sent in from infinity can come back out with a larger amplitude than it went in with,
provided the wave and the black hole satisfy certain conditions. This amplification phenomenon is known as superradiance. However, this phenomenon cannot be captured by solving the scattering problem using the WKB method, since $i\mathcal{K}$ is always real \cite{Konoplya2019}, and therefore Eq.~\eqref{GBF} yields positive GBFs.
% Evaluating the Wronskian of $\Psi$ and its complex conjugate from Eq.~\eqref{assym} near the horizon and again at infinity, and equating the two, yields a relation
% \begin{equation}
% 1 - |\mathcal{R}|^2 = \frac{\Omega - m\Omega_H}{\Omega} \, |\mathcal{T}|^2.
% \end{equation}
For a massless scalar field, when $0 < \Omega < m\,\Omega_H$, the reflected wave carries more energy than the incoming one and thus any mode whose frequency falls below $m\Omega_H$ is amplified rather than absorbed by the black hole. \\
% To quantify how much amplification a given mode experiences, we introduce the amplification factor
% \begin{equation}
% Z_{lm} \equiv |\mathcal{R}|^2 - 1 .
% \end{equation}

\noindent For the calculation of superradiance instability, we will restrict to
low-frequency modes $\Omega M \ll 1$ (which also implies $a\Omega \ll 1$). Under this condition, the matched asymptotics method applies: we split space into a near-horizon region, $r- r_+ \ll 1$, and a far region, $r - r_+\gg M$, solve the radial equation separately in each, and stitch the two solutions together in their common domain of validity \cite{Cardoso2004, Li2023, Yang2023}. First, we introduce the rescaled variable and parameters
\begin{equation}
z = \frac{r - r_+}{r_+ - r_-}, \qquad
\xi = \Omega(r_+ - r_-), \qquad
P = \frac{r_+^2 + a^2}{r_+ - r_-}\,(m\Omega_H - \Omega),
\end{equation}
where $r_+$ and $r_-$ are the horizons of the black hole.

\noindent $\bullet$ {\em Near-horizon region:} For $\Omega z \ll 1$, Eq.~\eqref{Radial} can be written in the following form
\begin{equation}
    z(z+1)\partial_z\left[z(z+1)\partial_z R(z)\right]+\left(P^2-z(z+1)A_{lm}\right)R(z)=0.
\end{equation}
Introducing a new variable $Q(z)=(z+1/z)^{-iP}R(z)$ reduces the equation to a hypergeometric form. We take the solution respecting the horizon boundary condition, for which we have
\begin{equation}
R = A_1 \left(\frac{z+1}{z}\right)^{iP} {}_2F_1(-l,\, l+1,\, 1-2iP,\, -z) .
\end{equation}
Here, $A_1$ is an integration constant, and we have considered $A_{lm}=l(l+1)$. The behaviour of $R(z)$ for large $z$ splits into two power-law pieces,
\begin{equation}
R(z) \sim A_1 z^{l}\,\frac{\Gamma(1-2iP)\Gamma(2l+1)}{\Gamma(1+l-2iP)\Gamma(l+1)}
+ A_1 z^{-l-1}\,\frac{\Gamma(1-2iP)\Gamma(-2l-1)}{\Gamma(-l)\Gamma(-l-2iP)} . \label{near}
\end{equation}
 
\noindent $\bullet$ {\em Far region:} As $z\to\infty$, the equation instead reduces to
\begin{equation}
\frac{d^2 R}{dz^2} + \frac{2}{z}\frac{dR}{dz}
+ \left[\xi^2 - \frac{l(l+1)}{z^2}\right] R(z) = 0,
\end{equation}
which is solved by confluent hypergeometric functions,
\begin{equation}
R(z) = e^{-i\xi z} C_1 z^{l} U(l+1, 2l+2, 2i\xi z)
+ e^{-i\xi z} C_2 z^{-l-1} U(-l, -2l, 2i\xi z) , \label{far}
\end{equation}
which for very small $z$ reduces to the same two power laws,
$R(z) \sim C_1 z^l + C_2 z^{-l-1}$. Matching the coefficients of $z^l$ and $z^{-l-1}$ between the two expansions (Eq.~\eqref{near} and Eq.~\eqref{far}) fixes $C_1$ and $C_2$ in terms of $A_1$. 

\noindent Finally, comparing the large - $r$ behavior of $R$ to the asymptotic form $R_\infty(r) \sim (A_I e^{-i\Omega r_*} + A_R e^{i\Omega r_*})/r$, gives the incoming and outgoing amplitudes explicitly in terms of $C_1$
and $C_2$:
\begin{align}
A_I &= C_1\,\frac{(-2i)^{-l-1}\xi^{-l}\Gamma(2l+2)}{\Omega \Gamma(l+1)}
+ C_2\,\frac{(-2i)^{l}\xi^{l+1}\Gamma(-2l)}{\Omega \Gamma(-l)}, \\
A_R &= C_1\,\frac{(2i)^{-l-1}\xi^{-l}\Gamma(2l+2)}{\Omega \Gamma(l+1)}
+ C_2\,\frac{(2i)^{l}\xi^{l+1}\Gamma(-2l)}{\Omega \Gamma(-l)} .
\end{align}
 
\noindent Combining these results, we get the amplification factor, which takes the closed form \cite{Li2023}
\begin{equation}
Z_{lm} = \frac{|A_R|^2}{|A_I|^2}-1 = 4P\,\xi^{2l+1}\,\frac{(l!)^4}{\big[(2l)!\big]^2\big[(2l+1)!!\big]^2}
\prod_{n=1}^{l} \left(1 + \frac{4P^2}{n^2}\right) ,
\end{equation}
valid for any spin $a \le M$ as long as $\omega M \ll 1$.

\begin{figure}[H]
    \centering

    \begin{subfigure}{0.48\textwidth}
        \centering
        \includegraphics[width=\linewidth]{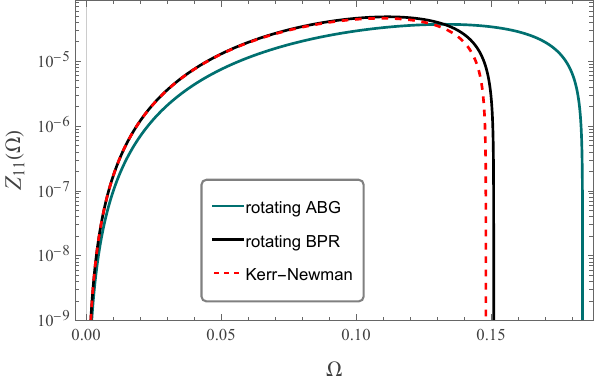}
    \end{subfigure}
    \hfill
    \begin{subfigure}{0.48\textwidth}
        \centering
        \includegraphics[width=\linewidth]{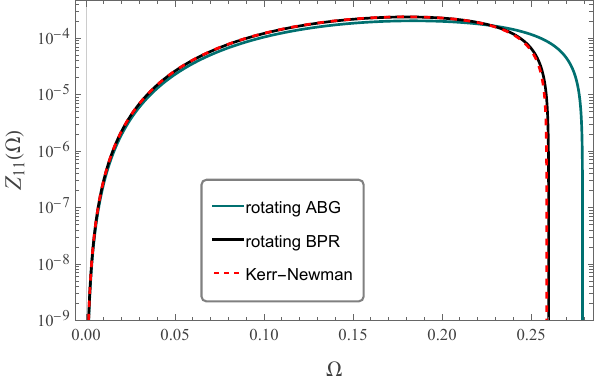}
    \end{subfigure}

    \caption{Superradiance amplification factors ($l=1$, $m=1$ and $M=1$) as a function of real $\Omega$ for rotating charged regular black holes. (left) For $a=0.5$, $q=0.4$. (right) For $a=0.8$, $q=0.2$.}
    \label{fig:SUP}
\end{figure}
 
\noindent Fig.~\ref{fig:SUP} shows $Z_{11}$ for two sets of the charge parameter $q$ and of the spin $a$. For massless scalar fields, the amplification starts with $\Omega>0$ and shuts off as $\Omega$ approaches the superradiant threshold $m\Omega_H$. The $\Omega$ dependence of the amplification factor for the Kerr--Newman black hole \cite{Xu2020fgq, Myung2022biw} and rotating BPR black hole closely follows one another. However, the amplification factor of the rotating ABG geometry is clearly distinct, with a slightly lower maximum value but with a higher threshold of $\Omega$. Larger spin boosts the amplification and extends the threshold. The charge parameter $q$ also influences the threshold frequency, where a larger $q$ pushes that threshold to somewhat higher $\Omega$ if the spin remains the same.

\section{Lyapunov exponent and shadow} \label{photon}

\noindent The photon sphere, the region of unstable circular null geodesics, governs several observable aspects of a spacetime. In rotating spacetimes, there is a split; photons co-rotating with the black hole have a smaller photon-sphere radius, while the counter-rotating photons orbit a larger radius. The photon sphere sets the boundary of the black hole shadow, since photons on these unstable orbits define the critical impact parameters separating captured from scattered light. It is also related to the ringdown signal in the eikonal (large multipole) limit. For rotating regular black holes, therefore, the study of unstable photon orbits provides a useful bridge between the underlying geometry, shadows, and perturbative dynamics. The photon trajectories can be obtained by studying the null geodesics of the rotating regular spacetime through the Hamilton--Jacobi (HJ) formalism \cite{Konoplya2018}. The HJ equation is given by
\begin{equation}
H(x^\mu,p_\mu)+\frac{\partial S}{\partial\tau}=0,
\label{eq:HJ_general}
\end{equation}
where $\tau$ denotes the affine parameter, $S$ is the Jacobi action, and $H$ represents the Hamiltonian associated with the null geodesic motion. For a stationary and axisymmetric spacetime, the metric is independent of the coordinates $t$ and $\phi$. Consequently, the energy $E$ and the azimuthal angular momentum $L$ are conserved quantities. In addition, symmetries of the spacetime give rise to another conserved quantity, namely the Carter constant \cite{Carter1968}, ensuring the separability of the HJ equation. We therefore adopt the following separable ansatz for the Jacobi action
\begin{equation}
S=-Et+L\phi+S_r(r)+S_\theta(\theta),
\label{eq:HJ_ansatz}
\end{equation}
\noindent where $E=-\partial S/\partial t$, $L=\partial S/\partial\phi$. Since the action in Eq.~\eqref{eq:HJ_ansatz} does not explicitly depend on the affine parameter, the HJ equation for null geodesics reduces to
\begin{equation}
H=\frac{1}{2}g^{\mu\nu}p_\mu p_\nu=0,
\label{eq:null_HJ}
\end{equation}
with canonical momenta $p_\mu=\partial S/\partial x^\mu$. Substitution of the separable action into Eq.~\eqref{eq:null_HJ} allows the HJ equation to be separated into independent radial and angular equations with associated potentials
\begin{equation}
    V_R(r)=\left((r^2+a^2)E-aL\right)^2-\Delta\left(C+(L-aE)^2\right)^2, \label{eq:radial}
\end{equation}
\begin{equation}
    V_\Theta(\theta)=C+\left(a^2E^2-L^2csc^2\theta\right)cos^2\theta,
\end{equation}
where $C$ denotes the Carter constant. The photon sphere is determined by the critical photon trajectories that separate photons falling into the black hole from those that are scattered back toward the asymptotic observer. Since the radial motion of photons is governed by the effective potential in Eq.~\eqref{eq:radial}, the critical trajectories correspond to unstable spherical photon orbits. These orbits satisfy the conditions
\begin{equation}
V_R(r_c)=0,
\qquad
V_R'(r_c)=0,
\qquad
V_R''(r_c)>0,
\label{eq:critical_conditions}
\end{equation}
where $r_c$ denotes the radius of the critical photon orbit and a prime indicates differentiation with respect to $r$. The first two conditions ensure that the photons remain on a circular orbit, while the third condition characterises the orbit's instability. 

\subsection{Lyapunov exponent}

\noindent The instability timescale of null geodesics trapped near the photon sphere of a spacetime is characterised by the Lyapunov exponent $\lambda$. For a stationary, axisymmetric case, the Lyapunov exponent governing the divergence of null rays near $r_{c}$ can be written in terms of the radial potential $V_R(r)$ and $\dot{t}$ (computed using the expressions of canonical momenta $p_t$ and $p_\phi$) as \cite{Cardoso2009}
\begin{equation}
    \lambda = \sqrt{\, -\,\frac{V_R''(r_{c})}{2\,(\dot{t})^2}} \, ,
    \qquad 
    \dot{t} 
    = \frac{1}{\Sigma(r_c)\Delta(r_c)}\left[E\{(a^2+r_c^2)^2-a^2\Delta(r_c)\}-2aLM(r_c)r_c\right] ,
    \label{eq:lyap-rot}
\end{equation}
which measures the exponential rate at which nearby null rays diverge from (or converge to) the unstable circular orbit. In the eikonal (large multipole number, $l \gg 1$) limit, this same photon-sphere quantity has a strong correlation with the black hole QNM spectrum. The imaginary part of the mode frequency, $\mathrm{Im}(\omega_{lmn}) = -(n+\tfrac{1}{2})|\lambda|$ (where $n=0,1,2,\dots$ is the overtone number), encodes the Lyapunov instability timescale of the photon sphere. Thus, the decay rate of high-$l$ QNMs is a geometric quantity. This geodesic--QNM correspondence was shown for a class of static, spherically symmetric geometries by Cardoso \emph{et al.} \cite{Cardoso2009} and was demonstrated for the Kerr black hole in \cite{Yang2012he}.

\begin{table}[H]
    \centering
    \footnotesize
    \setlength{\tabcolsep}{4pt}
    \renewcommand{\arraystretch}{1.2}
    \singlespacing
    \scalebox{1}{
    \begin{tabular}{|c|c||c|c|c||c|c|c||c|c|c|}
        \hline
         & & \multicolumn{3}{c||}{\textbf{rotating ABG}} & \multicolumn{3}{c||}{\textbf{rotating BPR}}
         & \multicolumn{3}{c|}{\textbf{Kerr--Newman}}\\
        \hline
        $q$ & $a$ & $\frac{1}{2}|\lambda|$ & -Im($\omega$) & $\delta$ & $\frac{1}{2}|\lambda|$ & -Im($\omega$) & $\delta$ & $\frac{1}{2}|\lambda|$ & -Im($\omega$) & $\delta$ \\
        \hline
        0.1 & 0.1 & $0.0960608$ & $0.0960692$ & $0.00874\%$ & $0.0961902$ & $0.0961981$ & $0.00821\%$ & $0.0961944$ & $0.0962023$ & $0.00821\%$ \\
            & 0.3 & $0.0951534$ & $0.0951633$ & $0.01040\%$ & $0.0953574$ & $0.0953654$ & $0.00839\%$ & $0.095364$ & $0.095372$ & $0.00839\%$ \\
            & 0.5 & $0.0925711$ & $0.0925842$ & $0.01415\%$ & $0.0929289$ & $0.0929378$ & $0.00958\%$ & $0.0929406$ & $0.0929494$ & $0.00947\%$ \\
            & 0.7 & $0.0859967$ & $0.0860155$ & $0.02186\%$ & $0.0867658$ & $0.0867772$ & $0.01314\%$ & $0.0867905$ & $0.0868017$ & $0.01290\%$ \\
            & 0.9 & $0.0640028$ & $0.0640278$ & $0.03906\%$ & $0.0670686$ & $0.0670839$ & $0.02281\%$ & $0.0671624$ & $0.0671773$ & $0.02233\%$ \\
        \hline
        0.3 & 0.1 & $0.0950936$ & $0.0951079$ & $0.01504\%$ & $0.0964553$ & $0.0964645$ & $0.00954\%$ & $0.0965631$ & $0.0965718$ & $0.00901\%$ \\
            & 0.3 & $0.0931457$ & $0.0931808$ & $0.03769\%$ & $0.0953841$ & $0.0953967$ & $0.01321\%$ & $0.0955556$ & $0.0955666$ & $0.01151\%$ \\
            & 0.5 & $0.0881445$ & $0.0882120$ & $0.07658\%$ & $0.0923967$ & $0.0924144$ & $0.01916\%$ & $0.0927032$ & $0.0927178$ & $0.01575\%$ \\
            & 0.7 & $0.0732604$ & $0.0733793$ & $0.16226\%$ & $0.0845778$ & $0.0846028$ & $0.02956\%$ & $0.085268$ & $0.0852882$ & $0.02369\%$ \\
        \hline
        0.5 & 0.1 & $0.0902968$ & $0.0903436$ & $0.05183\%$ & $0.0967203$ & $0.0967335$ & $0.01365\%$ & $0.0971748$ & $0.097186$ & $0.01153\%$ \\
            & 0.3 & $0.0823304$ & $0.0825231$ & $0.23410\%$ & $0.0949222$ & $0.0949486$ & $0.02781\%$ & $0.0956663$ & $0.0956865$ & $0.02112\%$ \\
            & 0.5 & $-$ & $-$ & $-$ & $0.0901285$ & $0.0901706$ & $0.04671\%$ & $0.0915418$ & $0.0915744$ & $0.03556\%$ \\
            & 0.7 & $-$ & $-$ & $-$ & $0.0758021$ & $0.0758536$ & $0.06794\%$ & $0.0796758$ & $0.0797244$ & $0.06101\%$ \\
        \hline
    \end{tabular}
    }
    \caption{Comparison between imaginary part of the fundamental modes for $l=m=20$ case and co-rotating Lyapunov exponents. Here $M=1$.}
    \label{tab:Lya_1}
\end{table}

\begin{table}[H]
    \centering
    \footnotesize
    \setlength{\tabcolsep}{4pt}
    \renewcommand{\arraystretch}{1.2}
    \singlespacing
    \scalebox{1}{
    \begin{tabular}{|c|c||c|c|c||c|c|c||c|c|c|}
        \hline
         & & \multicolumn{3}{c||}{\textbf{rotating ABG}} & \multicolumn{3}{c||}{\textbf{rotating BPR}}
         & \multicolumn{3}{c|}{\textbf{Kerr--Newman}}\\
        \hline
        $q$ & $a$ & $\frac{1}{2}|\lambda|$ & -Im($\omega$) & $\delta$ & $\frac{1}{2}|\lambda|$ & -Im($\omega$) & $\delta$ & $\frac{1}{2}|\lambda|$ & -Im($\omega$) & $\delta$ \\
        \hline
        0.1 & 0.1 & $0.0961262$ & $0.0961338$ & $0.00791\%$ & $0.0962139$ & $0.0962221$ & $0.00852\%$ & $0.0962168$ & $0.096204$ & $0.01330\%$ \\
            & 0.3 & $0.0957268$ & $0.0957341$ & $0.00763\%$ & $0.0957892$ & $0.0957977$ & $0.00887\%$ & $0.0957912$ & $0.0957998$ & $0.00898\%$ \\
            & 0.5 & $0.0950579$ & $0.0950558$ & $0.00221\%$ & $0.0951037$ & $0.0951066$ & $0.00305\%$ & $0.0951052$ & $0.0951083$ & $0.00326\%$ \\
            & 0.7 & $0.0942290$ & $0.0942198$ & $0.00976\%$ & $0.0942636$ & $0.0942605$ & $0.00329\%$ & $0.0942648$ & $0.0942618$ & $0.00318\%$ \\
            & 0.9 & $0.0933059$ & $0.0932854$ & $0.02197\%$ & $0.0933326$ & $0.0933199$ & $0.01361\%$ & $0.0933335$ & $0.0933211$ & $0.01329\%$ \\
        \hline
        0.3 & 0.1 & $0.0956897$ & $0.0956906$ & $0.00094\%$ & $0.0965903$ & $0.0965971$ & $0.00704\%$ & $0.096663$ & $0.0966702$ & $0.00745\%$ \\
            & 0.3 & $0.0955883$ & $0.0955790$ & $0.00973\%$ & $0.0962178$ & $0.0962227$ & $0.00509\%$ & $0.0962692$ & $0.0962753$ & $0.00634\%$ \\
            & 0.5 & $0.0950968$ & $0.0950288$ & $0.07151\%$ & $0.0955545$ & $0.0955415$ & $0.01360\%$ & $0.0955922$ & $0.0955838$ & $0.00879\%$ \\
            & 0.7 & $0.0943775$ & $0.0942847$ & $0.09831\%$ & $0.0947203$ & $0.0946961$ & $0.02555\%$ & $0.0947487$ & $0.0947303$ & $0.01942\%$ \\
        \hline
        0.3 & 0.1 & $0.0933110$ & $0.0932854$ & $0.02744\%$ & $0.0971926$ & $0.0971952$ & $0.00268\%$ & $0.0974944$ & $0.0974988$ & $0.00451\%$ \\
            & 0.3 & $0.0944164$ & $0.0943482$ & $0.07223\%$ & $0.0969848$ & $0.0969790$ & $0.00598\%$ & $0.0971964$ & $0.0971954$ & $0.00103\%$ \\
            & 0.5 & $-$ & $-$ & $-$ & $0.0964002$ & $0.0963399$ & $0.06255\%$ & $0.0965545$ & $0.0965154$ & $0.04050\%$ \\
            & 0.7 & $-$ & $-$ & $-$ & $0.0955988$ & $0.0955121$ & $0.09069\%$ & $0.0957145$ & $0.0956554$ & $0.06175\%$ \\
        \hline
    \end{tabular}
    }
    \caption{Comparison between imaginary part of the fundamental modes for $l=20$, $m=-20$ case and counter-rotating Lyapunov exponents. Here $M=1$.}
    \label{tab:Lya_2}
\end{table}

\noindent The results presented in Tables~\ref{tab:Lya_1} and \ref{tab:Lya_2} demonstrate excellent correspondence between the Lyapunov exponents and the imaginary part of the fundamental QNM mode for all three spacetimes (rotating ABG, rotating BPR, Kerr--Newman). The relative discrepancies ($\delta$) remain below $0.3\%$ for the parameter space explored in the tables. 

\subsection{Shadow}

\noindent A critical observational signature of a black hole or black hole mimicker is its shadow profile.  In order to get the shadow profile, we need the critical parameters, which can be obtained from Eq.~\eqref{eq:critical_conditions} and can be written as
\begin{equation}
\chi = \frac{L}{E} =
\frac{
a^2+r_c^2
-(4r_c\Delta(r_c))/\Delta'(r_c)
}{a},
\end{equation}
\begin{equation}
\eta = \frac{C}{E^2} =
\frac{r_c^2 \left[16 a^2 \Delta \left(r_c\right)-\left(r_c \Delta '\left(r_c\right)-4 \Delta \left(r_c\right)\right){}^2\right]}{a^2 \Delta '\left(r_c\right){}^2}.
\end{equation}
These quantities determine the family of critical null geodesics and, consequently, the boundary of the shadow. However, an observer at a large distance does not directly observe the radial coordinates of these photon orbits. Instead, the observer sees their projection onto the celestial plane, which is perpendicular to the line joining the observer and the black hole \cite{dewitt1973black}. At the asymptotic limit $r\rightarrow\infty$, the celestial coordinates ($\alpha,\beta$) can be expressed in terms of the conserved quantities as
\begin{equation}
\alpha(r_c)=-\frac{1}{\sin\theta_0}\chi, \qquad
\beta^2(r_c)=\eta-\chi^2\cot^2\theta_0+a^2\cos^2\theta_0,
\label{eq:alphabeta}
\end{equation}
where $\theta_0$ is the inclination angle between the observer's line of sight and the black hole rotation axis. Thus, the parametric curve ${\alpha(r_c),\beta(r_c)}$ provides the apparent shadow profile of the rotating regular black hole. The resulting shadow depends on the inclination angle $\theta_0$, the rotation parameter $a$, and the parameters entering the mass function $M(r)$.
\begin{figure}[H]
    \centering

    \begin{subfigure}{0.32\textwidth}
        \centering
        \includegraphics[width=\linewidth]{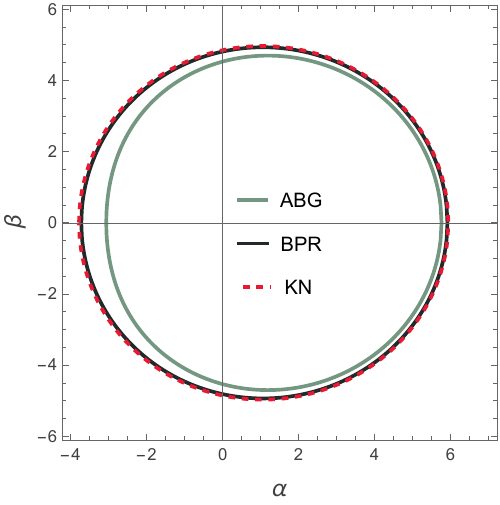}
    \end{subfigure}
    \hfill
    \begin{subfigure}{0.32\textwidth}
        \centering
        \includegraphics[width=\linewidth]{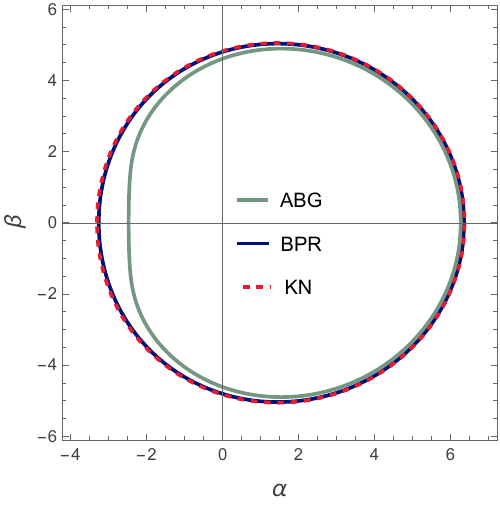}
    \end{subfigure}
    \hfill
    \begin{subfigure}{0.32\textwidth}
        \centering
        \includegraphics[width=\linewidth]{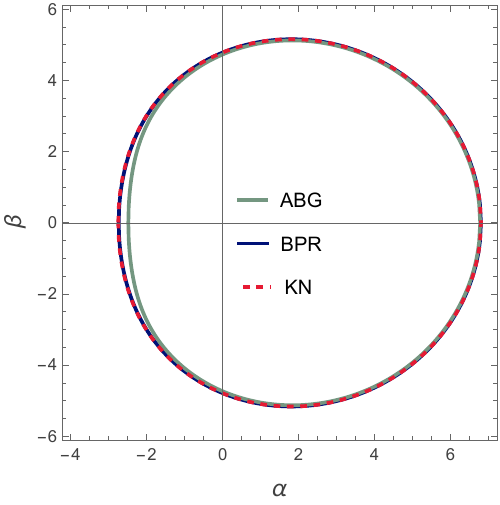}
    \end{subfigure}

    \caption{Shadow of Kerr--Newman and rotating regular black holes for (left) $a=0.5M$, $q=0.5M$, (middle) $a=0.7M$, $q=0.4M$, (right) $a=0.9M$, $q=0.2M$ at an inclination angle $\theta_0=90^\circ$.}
    \label{fig:shadow}
\end{figure}
\noindent From Fig.~\ref{fig:shadow}, one can observe that the shadow profiles of rotating BPR and Kerr--Newman black hole \cite{Tsukamoto2017fxq, Xavier2020egv, Sui2023rfh} almost overlap, while that of the rotating ABG metric \cite{Ban2026} is clearly distinct. A similar feature was also observed in the QNM spectra and GBFs (QNMs and GBFs of rotating BPR and Kerr--Newman black holes were very close, while that of rotating ABG metric differed significantly).\\

\begin{figure}[H]
    \centering

    \begin{subfigure}{0.48\textwidth}
        \centering
        \includegraphics[width=\linewidth]{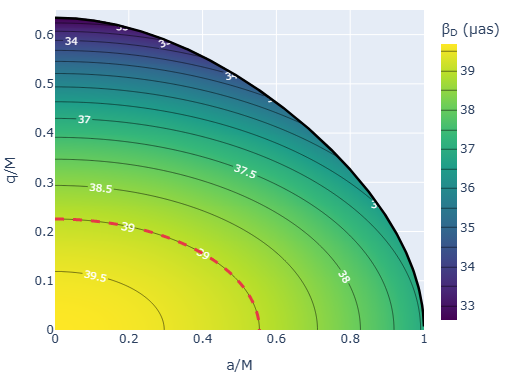}
    \end{subfigure}
    \hfill
    \begin{subfigure}{0.48\textwidth}
        \centering
        \includegraphics[width=\linewidth]{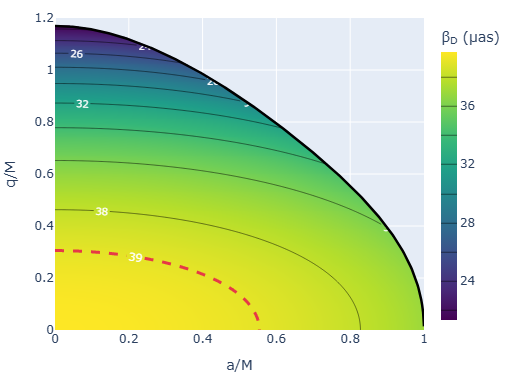}
    \end{subfigure}

    \caption{Density plot of angular shadow diameter $\beta_D$ with inclination angle $\theta_0=17^\circ$. (left) rotating ABG black hole. (right) rotating BPR black hole. The lower limit of $\beta_D$ for M87$^{*}$ is marked (red dashed line). }
    \label{fig:M87}
\end{figure}

\begin{figure}[H]
    \centering

    \begin{subfigure}{0.48\textwidth}
        \centering
        \includegraphics[width=\linewidth]{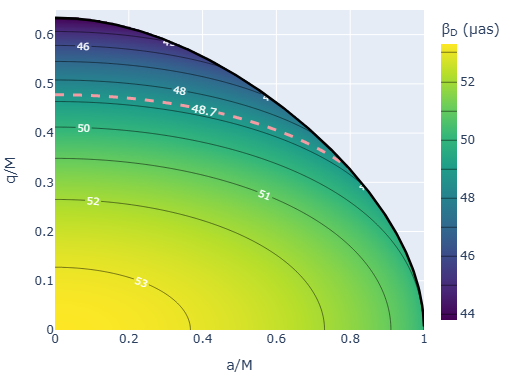}
    \end{subfigure}
    \hfill
    \begin{subfigure}{0.48\textwidth}
        \centering
        \includegraphics[width=\linewidth]{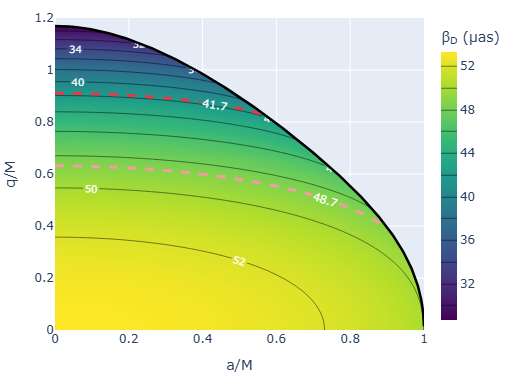}
    \end{subfigure}

    \caption{Density plot of angular shadow diameter $\beta_D$ with inclination angle $\theta_0=50^\circ$. (left) rotating ABG black hole. (right) rotating BPR black hole. The nominal value (dashed pink) and the lower limit (dashed red) of $\beta_D$ for Sgr A$^{*}$ are shown. }
    \label{fig:SgrA}
\end{figure}

\noindent Having obtained the shadow profiles, we can ask whether we can put bounds on metric parameters of the rotating regular black holes based on EHT observations. For M87$^{*}$, the EHT 2019 results give a measured angular shadow diameter $\beta_D = 42 \pm 3$ $\mu$as \cite{M872019}, with independently determined mass $M = 6.5 \times 10^9 M_\odot$ and distance $D = 16.8$ Mpc (from stellar population measurement \cite{Blakeslee2009, Bird2010, Cantiello2018}). For Sgr A$^{*}$, EHT 2022 results give $\beta_D=48.7 \pm 7$ $\mu$as \cite{EventHorizonTelescopeCollaboration2022}, with mass $M = 4.3 \times 10^6 M_\odot$, and distance $D = 8.277$ kpc (Gravity collaboration \cite{GRAVITY2021xju}). We will use the above values of mass and distance to compute $\beta_D$ for the rotating ABG and rotating BPR spacetimes using
\begin{equation}
    \beta_D=2\frac{R_{sh}}{D}=\frac{2}{D}\sqrt{\frac{A_{sh}}{\pi}},
\end{equation}
where $R_{sh}$ is the average shadow radius calculated from the area enclosed by the shadow profile $A_{sh}$. When comparing with the M87$^{*}$, we set the inclination angle to $\theta_0=17^\circ$ and identify the parameter space for which $39 < \beta_D < 45$ $\mu$as (Fig. \ref{fig:M87}). For Sgr A$^{*}$, we set $\theta_0=50^\circ$ and search for values of $a/M$ and $q/M$ corresponding to $41.7 < \beta_D < 55.7$ $\mu$as (Fig. \ref{fig:SgrA}). Fig \ref{fig:M87} and \ref{fig:SgrA} show that the angular shadow diameter decreases with increasing spin and charge parameter. The observational bounds (lower limit of $\beta_D$) are marked with dashed red lines. Thus, to be compatible with EHT data, the values of $a/M$ and $q/M$ have to be below the red dashed lines. The M87$^{*}$ shadow gives a stricter bound on spin and charge parameter, while for Sgr A$^{*}$, the entire parameter space of rotating ABG black hole and a significant part of the rotating BPR parameter space lies within the allowed region. The bounds on $a/M$ and $q/M$ of rotating ABG (close to \cite{Ban2026} for M87$^{*}$) and rotating BPR black holes are listed in Table \ref{tab:shadow_comparison}. For Sgr A$^{*}$, we could not put any bounds on rotating ABG spacetime based on the shadow diameter, and the corresponding bounds in the table simply ensure the existence of horizons.

\begin{table}[ht]
\centering

\renewcommand{\arraystretch}{1.5}

\resizebox{\textwidth}{!}{%
\begin{tabular}{lllllll}
\hline
\noalign{\vskip 1mm}
\textbf{Black hole} &
$M$ ($M_\odot$) &
\quad$D$ (kpc) &
\quad$\theta_0$ &
\quad$\beta_D$ ($\mu$as) &
\textbf{\qquad rotating ABG constraint} &
\textbf{\qquad rotating BPR constraint} \\
\noalign{\vskip 1mm}
\hline
\noalign{\vskip 1mm}

M87$^{*}$ &
$6.5\times10^9$ &
\quad$16800$ &
\quad$17^\circ$ &
\quad$39 < \beta_D < 45$ &
\qquad $a/M \lesssim 0.554$ \quad $q/M\lesssim0.225$ &
\qquad $a/M \lesssim 0.553$ \quad $q/M\lesssim0.306$ \\

Sgr A$^{*}$ &
$4.3\times10^6$ &
\quad$8.277$ &
\quad$50^\circ$ &
\quad$41.7 < \beta_D < 55.7$ &
\qquad $a/M < 1$ \qquad \hspace{1.4mm} $q/M\lesssim0.634$ &
\qquad $a/M < 1$ \qquad \hspace{1.4mm} $q/M\lesssim0.912$ \\

\noalign{\vskip 1mm}
\hline
\end{tabular}%
}

\caption{Bound on the parameter space of rotating ABG and BPR black holes from angular shadow diameter.}
\label{tab:shadow_comparison}
\end{table}

\section{Conclusion} \label{conclusion}

\noindent In this article, we have constructed charged rotating regular black holes corresponding to the ABG and BPR mass function using the Newman-Janis algorithm and carried out a detailed comparison of their geometric, perturbative and scattering properties.\\

\noindent The ergo-region, bounded by the static limit surface, grows with increasing spin in the manner familiar from Kerr and Kerr--Newman. Using a minimal set of Weyl, Ricci, and mixed curvature invariants, we confirmed that the rotating geometries remain regular everywhere for non-zero charge, while the associated energy-momentum tensor was found to violate the weak energy condition over a finite range of the radial coordinate.\\

\noindent Scalar QNMs, computed independently via the spectral method and the sixth-order WKB method, show a consistent qualitative pattern across all three geometries: increasing spin or charge shifts the real part of the frequency to larger values (for positive values of $m$) while decreasing the magnitude of the imaginary part, and rotation breaks the mirror symmetry of the spectrum that is present in the static limit. For negative values of azimuthal number ($m$), we observe a qualitative difference in charge parameter dependence between rotating ABG and rotating BPR/Kerr--Newman black holes.\\

\noindent The QNM-GBF correspondence links the QNM spectrum to the scattering problem and is verified for the two rotating regular black holes we studied. The low-frequency superradiant amplification factor, derived analytically via a matched asymptotic expansion, shows that amplification and frequency threshold are enhanced by both spin and charge.\\

\noindent The photon sphere and associated shadow boundary are obtained from the Hamilton-Jacobi formalism, and their correspondence to the QNM damping rate via the Lyapunov exponent of the unstable photon orbit is shown. We have also put bounds on the parameter space of rotating ABG and rotating BPR black holes from EHT observations.\\

\noindent A recurring and notable result across every probe considered is that the rotating BPR and Kerr--Newman black holes remain nearly indistinguishable from one another, while the rotating ABG geometry consistently stands apart, particularly at larger values of spin and charge. This suggests that the observable signatures of rotating regular black holes are not uniquely determined by the requirements of regularity and asymptotic Kerr--Newman behaviour alone, but remain sensitive to the specific form of the mass function.\\

\noindent A crucial next step will be to extend the analysis to gravitational perturbations, which couple more directly to potential gravitational wave observations; assessing whether the rotating charged regular black holes can be observationally distinguished from one another and from the classical Kerr/Kerr--Newman black hole, given realistic future detector sensitivities. It is also important to examine whether the type-I Newman-Janis construction used here, as opposed to alternative prescriptions for complexification, affects the degree of degeneracy observed among different black hole geometries.

\noindent \section*{ACKNOWLEDGEMENTS}

\noindent ABM is grateful to Prof. Sayan Kar for his continued guidance and acknowledges his thorough reading of this manuscript. He thanks Shailesh Kumar and Anjan Kar for their valuable feedback. He also thanks the Indian Institute of Technology Kharagpur, India, for supporting
him through a fellowship.

\appendix

\section{Other charged regular black holes} \label{appendix}

\noindent We have examined the astrophysical signatures of two charged rotating regular black holes, namely the rotating ABG and the rotating BPR black hole. However, there are many more regular spacetimes that approach Reissner--Nordstr\"om geometry at the large-r limit. The mass functions of a few such regular geometries are listed in Table~\ref{tab:regular_bh}. Plugging in these mass functions in Eq.~\eqref{regrot}, we obtain the rotating counterparts of these regular black holes. Now, we will compare the QNM spectra of these rotating geometries with that of the Kerr--Newman black hole. Table~\ref{tab:qnm_reg} shows the fundamental modes of different charged rotating regular black holes along with Kerr--Newman spacetime for $q=0.3M$ and $l=2$, $m=2$ case.\\

\begin{table}[H]
\centering
\begin{tabular}{@{}l l@{}}
\hline
\noalign{\vskip 1mm}
\textbf{Black hole} & \textbf{\qquad Mass function} \\
\noalign{\vskip 1mm}
\hline\\
Dymnikova \cite{Dymnikova2004} &
$\displaystyle \qquad M(r)=\frac{2M}{\pi}\left(\arctan(r/r_0)-\frac{r r_0}{r^2+r_0^2}\right)$, \quad $r_0=\frac{\pi q^2}{8M}$  \\[10pt]

Ay\'on-Beato--Garc\'ia (ABG II) \cite{AynBeato1999}  &
$\displaystyle \qquad M(r)=M\left(1-\tanh\left(\frac{q^2}{2Mr}\right)\right)$ \\[10pt]

Balart--Vagenas (BV I) \cite{Balart2014} &
$\displaystyle \qquad M(r)=M\exp\left(-\frac{q^2}{2Mr}\right)$ \\[10pt]

Balart--Vagenas (BV II) \cite{Balart2014jia} &
$\displaystyle \qquad M(r)=M\left[1-\left(1+\left(\frac{2Mr}{q^2}\right)^3\right)^{-1/3}\right]$ \\[16pt]

\hline
\end{tabular}
\caption{Mass functions for different charged regular black holes.}
\label{tab:regular_bh}
\end{table}

\noindent One can observe that the values of QNM frequencies of the rotating regular spacetimes are very close to the QNM spectra of the Kerr--Newman black hole. Thus, these spacetimes are almost indistinguishable from Kerr-Newman geometry using the ringdown signal. Owing to the QNM-GBF correspondence, the GBFs for these spacetimes will also be nearly identical. Though not shown here, the shadow profiles of these geometries virtually overlap. Thus, among the charged rotating regular black holes studied in this work, the rotating ABG black hole stands out in all astrophysical observables, while the rotating BPR black hole might be distinguishable for large values of the charge parameter.

\begin{table}[H]
    \centering
    \footnotesize
    \setlength{\tabcolsep}{4pt}
    \renewcommand{\arraystretch}{1.2}
    \singlespacing
    \scalebox{1}{
    \begin{tabular}{|c|c|c|c|c|c|}
        \hline
        $a$ & Kerr--Newman & rotating Dymnikova & rotating ABG II & rotating BV I & rotating BV II\\
        \hline
        0.1 & $0.507806 - 0.0970886i$ & $0.507803 - 0.097083i$   & $0.507804 - 0.0970843i$ & $0.507667 - 0.0969009i$ & $0.507804 - 0.0970826i$ \\
        0.3 & $0.547467 - 0.0960848i$ & $0.547473 - 0.0960805i$  & $0.547471 - 0.0960815i$ & $0.547615 - 0.0959449i$ & $0.547474 - 0.09608i$ \\
        0.5 & $0.600158 - 0.0932893i$ & $0.600164 - 0.0932957i$  & $0.600163 - 0.0932943i$ & $0.600267 - 0.093486i$  & $0.600166 - 0.0932949i$ \\
        0.7 & $0.678031 - 0.0860895i$ & $0.678021 - 0.0861066i$  & $0.678023 - 0.0861027i$ & $0.677611 - 0.0865773i$ & $0.678025 - 0.0861047i$ \\
        0.9 & $0.835755 - 0.0587944i$ & $0.83568 - 0.0588001i$   & $0.835696 - 0.0587993i$ & $0.833659 - 0.0589339i$ & $0.835704 - 0.0587887$ \\
        \hline
    \end{tabular}}
    \caption{QNM fundamental modes for $l=m=2$ case. Here $M=1$ and $q=0.3$.}
    \label{tab:qnm_reg}
\end{table}

\section{Methods for computing QNMs} \label{methods}
\noindent There are a plethora of analytic, semi-analytic and numerical schemes that can be utilised to obtain the QNM spectra. Here, we briefly discuss the two methods we have employed to compute the QNM frequencies of rotating charged regular black holes, starting with the spectral method.\\

% \subsubsection{Spectral method}

\noindent $\bullet$ {\em Spectral method :} This is one of the numerical techniques employed to solve eigenvalue problems arising from differential equations. For this, one needs to expand the perturbation equations in a finite series of global basis functions -- Chebyshev polynomials being a very common choice because of their excellent approximation properties and the availability of well-conditioned differentiation matrices. First, we need to choose proper boundary conditions for the radial equation, ensuring a purely ingoing wave at the horizon and a purely outgoing wave at spatial infinity. For that, we examine the behaviour of $R(r)$ near the horizon and at infinity and define a new variable $X(r)$ using a suitable ansatz:
\begin{equation}
    R(r)=\exp(i \omega r)\left(\frac{r-r_+}{r-r_-}\right)^{-i \sigma}\left(\frac{r-\bar r_+}{r-\bar r_-}\right)^{i \sigma}r^{2iM\omega}X(r)
\end{equation}
where  $\sigma=(2r_+ M(r_+)\omega-a m)/\Delta'(r_+)$, with $r_+$ and $r_-$ being the event horizon and inner horizon of the black hole, respectively, while $\bar r_+$ and $\bar r_-$ are the negative roots of the equation $\Delta(r)=0$ (if there are no negative roots, we set $\bar r_+$ and $\bar r_-$ to zero). $M$ is the ADM mass of the geometry. Now, we compactify the domain, mapping the semi-infinite range from horizon ($r_+$) to spatial infinity onto a finite interval $[-1,1]$, suitable for the Chebyshev grid, by introducing a new coordinate $x=1-2r_+/r$. Next, the function $X(x)$ is to be discretised. Following \cite{jansen2017}, we expand $X$ in a basis of Chebyshev polynomials of the first kind, $C_i(x)$, evaluated at the collocation points,
\begin{equation}
    X(x) \approx \sum_{i=0}^{N} X(x_i)\, C_i(x),
\end{equation}
where $x_i$ are the Chebyshev--Gauss (roots) collocation points, $x_i = \cos\!\left(\frac{(2i+1)\pi}{2(N+1)}\right)$, ($i = 0,1,2,\dots,N$).

\noindent The radial equation now takes the form
\begin{equation}
    c_2(x_i,\omega)\,X''(x_i) + c_1(x_i,\omega)\,X'(x_i) + c_0(x_i,\omega)\,X(x_i) = 0, \label{sp1}
\end{equation}
where primes denote derivatives with respect to $x$. Each coefficient function $c_k(x,\omega)$, ($k=0,1,2$), is in general a polynomial in the (complex) quasinormal frequency $\omega$, which we expand as $c_k(x,\omega) = \sum_{a=0}^{p} c_{a,k}(x)\,\omega^{a}$. Thus $c_{a,k}(x)$ denotes the coefficient of $\omega^{a}$ appearing in $c_k(x,\omega)$, and $p$ is the highest power of $\omega$ occurring in Eq.~\eqref{sp1}. Collecting all the terms proportional to $\omega^{a}$ we define the matrix
\begin{equation}
    (M_a)_{ij} = c_{a,0}\,D^{(0)}_{ij} + c_{a,1}\,D^{(1)}_{ij} + c_{a,2}\,D^{(2)}_{ij}, \qquad a = 0,1,\dots,p, \label{c_aa}
\end{equation}
where, $D^{(n)}_{ij}$ denotes n-th derivative of $C_i(x_j)$, with $D^{(0)}_{ij} = \delta_{ij}$ the identity matrix. The discretised radial equation \eqref{sp1} is thus recast as the matrix polynomial eigenvalue problem
\begin{equation}
    \left(M_0 + \omega M_1 + \omega^2 M_2 + \cdots + \omega^{p} M_p\right)X = 0. \label{sp2}
\end{equation}
We need to solve this matrix equation to get $\omega$. For the angular function $S(\theta)$, we can choose the ansatz
\begin{equation}
    S(u)=\exp(a \omega u)(1+u)^{|m|/2}(1-u)^{|m|/2}Y(u)
\end{equation}
where $u=cos(\theta)$. For solving the angular equation, the same prescription detailed above is repeated, with $\omega$ replaced by the separation constant $A_{lm}$. We have used $N=40$ for the radial equation and $N=20$ for the angular equation.\\

% \subsubsection{WKB method}

\noindent $\bullet$ {\em WKB method :} An alternative, semi-analytic approach to computing QNM frequencies is the Wentzel--Kramers--Brillouin approximation, first applied to black hole perturbation theory by Schutz and Will \cite{Schutz1985}. For the WKB method, we need to express the radial equation in the following form
\begin{equation}
    \frac{d^2 \Psi}{d r_*^2} + Q(r) \Psi(r) = 0.
\end{equation}
We can rewrite Eq.~\eqref{Radial} in this form by defining a new variable $\Psi(r)=\sqrt{r^2+a^2}R(r)$
\begin{equation}
    \frac{d^2\Psi}{dr_*^2}+\left(\frac{a^2 m^2+\omega ^2 \left(a^2+r^2\right)^2-\Delta \left(a^2 \omega ^2+A_{lm} \right)-4 a m \omega  r M(r)}{\left(a^2+r^2\right)^2}-\frac{\Delta \left(\left(a^2-2 r^2\right) \Delta+r \left(a^2+r^2\right) \Delta'\right)}{\left(a^2+r^2\right)^4}\right)\Psi=0, \label{rad_wkb}
\end{equation}
where $dr_*/dr=\Delta/(r^2+a^2)$. The aim is to match an asymptotic WKB expansion of $\Psi$ in the two regions on either side of the potential peak to a Taylor expansion of $Q(r_*)$ around the maximum of the potential $r_*^{(0)}$, where $Q'(r_*^{(0)})=0$. Matching the solutions and their derivatives across the turning points using standard WKB connection formulas yields a condition on $\omega$
\begin{equation}
    \frac{iQ_0}{\sqrt{2 Q_0''}} -\sum_{i=0}^{k} \Lambda_i= \left(n+\frac{1}{2}\right) \qquad n = 0,1,2,\dots, \label{wkb2}
\end{equation}
where $Q_0 \equiv Q(r_*^{(0)})$ , $Q_0'' $ is the second derivative of $Q$ at $r_*^{(0)}$, and $n$ is the overtone number, with $n=0$ corresponding to the fundamental mode. $\Lambda_i$ are the WKB correction terms, and can be found in \cite{Iyer1987} (3rd order),\cite{Konoplya2003} (6th order), \cite{Matyjasek2017} (13th order). However, in our case, a complication arises as $Q$ is also a function of $\omega$ (thus $r_*^{(0)}$ also depends on $\omega$). To tackle this problem, we first solve $Q'=0$ for the $a=0$ case to compute $r_*^{(0)}$, and subsequently compute $\omega$ from Eq.~\eqref{wkb2}. Now, starting from the values of $r_*^{(0)}$ and $\omega$ obtained at $a=0$, following \cite{Tang2025}, we increment the spin parameter in small steps and, at each step, numerically solve Eq.~\eqref{wkb2} and $Q'(r_*^{(0)})=0$ simultaneously for $r_*^{(0)}$ and $\omega$, taking the solution from the previous spin value as the initial guess for the next.

\bibliographystyle{apsrev4-2}       % APS-like style for physics # spphys
\bibliography{ref}

\end{document}